\documentclass[12pt]{iopjournal}

\usepackage{amsmath, amsfonts, multirow}
\usepackage{lmodern}
\usepackage{float}
\usepackage{ragged2e}   
\usepackage{microtype}  
\usepackage{makecell} 

\microtypesetup{
  activate = true,
  expansion = true,
  protrusion = true
}
\usepackage[numbers,square]{natbib}
\AtBeginDocument{\justifying}

\usepackage{etoolbox}
\AtBeginEnvironment{abstract}{\justifying}

\AtBeginDocument{%
  \pagestyle{plain}
  \thispagestyle{plain}
}

\usepackage{sectsty}
\sectionfont{\Large\bfseries}         
\subsectionfont{\large\bfseries}      
\subsubsectionfont{\normalsize\bfseries}

\usepackage{caption}
\begin{document}


\title{\LARGE Sluggish random walks with resetting: nonequilibrium aspects}

\author{Theophilus Dwamena Frimpong$^{1,2}$, L\'ea Bresque$^1$ and \'Edgar Rold\'an$^{1,*}$}

\affil{$^1$ICTP—The Abdus Salam International Centre for Theoretical Physics, Strada Costiera 11, 34151, Trieste, Italy}

\affil{$^2$The University of Texas Rio Grande Valley, 1201 W. University Dr., 78539, Edinburg, TX, USA}





\begin{abstract}
We derive exact analytical expressions for the fixed-time probability distribution and the Laplace transform of the distribution of the first-passage time to reach a target for a sluggish random walk in one dimension subject to two different types of stochastic resetting: stochastic resetting at random Poisson-distributed times, and resetting implemented whenever a position threshold is reached. The optimal resetting rate that minimises the mean first-passage time is obtained as a function of the sluggishness parameter, while under threshold resetting the mean first-passage time to the origin is found to increase monotonically with the threshold. We also characterise the occupation-time statistics, retrieving a Lamperti-type distribution under free dynamics and under Poissonian resetting at rate for short observation times, a Gaussian distribution centred at half for long observation times, and a distinct deterministic limit under threshold resetting. To further characterise nonequilibrium aspects of resetting, we evaluate the distribution and mean of the cost incurred per reset under Poissonian resetting, and the cost rate under threshold resetting, for different (linear, power-law, exponential) cost functions, and find the same critical behaviour of the cost rate in the limits of a vanishing resetting rate and an infinite threshold, respectively. Numerical simulations validate with excellent accuracy our analytical formulae.
\end{abstract}

\section{Introduction}
Random walks are paradigmatic stochastic processes in physics comprised of successive steps of random nature~\citep{Hughes_1998} and serve as the mathematical basis for understanding diffusive behaviour in various scientific fields~\citep{Einstein_1905, Masuda_2017, Kutner_2017}, for instance, Brownian motion. Modern theoretical developments have extended the classical random-walk theory to include anomalous diffusion phenomena~\citep{metzler2000random, metzler2004restaurant, Bouchaud_1990,sposini2022towards}. Unlike normal diffusion, where the mean squared displacement grows linearly in time as~$\langle x^2(t)\rangle \sim t$, anomalous diffusion is characterised by a power-law scaling $\langle x^2(t)\rangle \sim t^\gamma$, where $\gamma \neq 1$~\citep{metzler2014anomalous}. The cases $\gamma < 1$ and $\gamma > 1$ correspond respectively to subdiffusion, in which the particle spreads more slowly than a Brownian walker, and superdiffusion, in which it spreads faster~\citep{metzler2000random}. 

Particularly interesting are the recently introduced sluggish random walks~\citep{zodage2023sluggish} where the walker's mobility and associated diffusion coefficient systematically decrease with increasing distance from the origin, inducing a subdiffusive behaviour that deviates from classical Brownian scaling laws~\cite{metzler2000random}. Such sluggish behaviour is motivated, for instance, by trap models of slow dynamics~\citep{Bouchaud_1992, Monthus_1996}, in which a particle becomes increasingly trapped as it explores regions further from the origin.

In parallel to the rapid development of anomalous diffusion, stochastic resetting has emerged as an active area of research in physics of stochastic phenomena. Put simply, stochastic resetting models a mechanism by which a stochastic (e.g., diffusive) process undergoes a sudden transformation (e.g., reset to the initial position) at a random time. Such resetting mechanism yields, for specific parameter values, an improved efficiency of diffusive exploration to find a target region~\citep{evans2011diffusion, Evans_2020}. The resetting (or restart) process prevents a walker from moving too far and transforms unrestricted exploration into a more controlled search process with finite mean search durations~\citep{Reuveni_2016}. The mathematical theory of stochastic resetting has explored various reset strategies, such as resetting to the maximum position~\citep{majumdar2015random} and power-law distributed reset times~\citep{nagar2016diffusion}. 

Diffusion with a position-dependent diffusivity has been combined with Poissonian resetting in several studies~\citep{Lenzi_2022, Menon_2024}. For sluggish walkers specifically, resetting has been considered for the initial position~\citep{DelVecchio_2026} and, very recently, for resetting to previously visited positions~\citep{Boyer_2026}. To our knowledge, a systematic treatment of the sluggish walk under Poissonian and threshold resetting, covering stationary states, first-passage and occupation-time statistics, and the cost of resetting, is still lacking. This gap is significant, as in many physical and biological systems anomalous subdiffusion and resetting-like mechanisms coexist. In this paper, we address this gap by developing an exact analytical framework for sluggish random walks subject to Poissonian~\citep{evans2011diffusion} and threshold-based resetting~\citep{Pinsky_2020, Bhat_2016}, and characterise some of their nonequilibrium properties. Our results reveal how the sluggishness parameter $\alpha$ introduced below qualitatively reshapes the effects of resetting in ways that are absent in normal diffusion.  

The rest of the paper is organised as follows. Section~\ref{sec:2} introduces the key building blocks, namely the sluggish random walk and two paradigmatic models of diffusion with stochastic resetting. In Section~\ref{sec:3} we build on these foundations to propose our models of sluggish random walks with each resetting protocol and derive the governing Fokker--Planck equation for the Poissonian case and the corresponding equation with a nonlocal boundary condition for the threshold case. Exact closed-form expressions for the steady-state position distributions are then derived in Section~\ref{sec:4}, a modified Bessel function representation for the Poissonian case and a Beta-type distribution for the threshold case obtained through a renewal argument, both tested with numerical simulations. Section~\ref{sec:5} discusses in depth first-passage times, deriving the mean first-passage time analytically for the two resetting protocols that we consider, together with the optimal resetting rate in the Poissonian case and the monotonic dependence on the threshold in the threshold case. Section~\ref{sec:6} characterises the occupation statistics of the walker, obtaining the full occupation time distribution via its double Laplace transform for Poissonian resetting and its deterministic long-time limit for threshold resetting. The cost of sustaining each nonequilibrium stationary state is examined in Section~\ref{sec:7}, where we derive the so-called resetting entropy production rate and the statistics of the cost per reset event, and identify the critical scaling of the mean cost under Poissonian resetting and, for large thresholds, under threshold resetting. Section~\ref{sec:8} closes with a summary of our main findings and an outlook on open questions.

\section{Preliminaries: sluggish random walks, and diffusion with resetting}
\label{sec:2}
\smallskip
Before developing our results on the statistics of sluggish diffusions with resetting, we first give a didactic review of sluggish diffusions (Sec.~\ref{sec21}) and ordinary diffusion with resetting (Poissonian in Sec.~\ref{sec22} and threshold-based in Sec.~\ref{sec23}).

\subsection{Sluggish random walks}
\label{sec21}
\smallskip
Unlike ordinary random walks, which are characterised by constant, space-homogeneous hopping probabilities, sluggish random walks are characterised by position-dependent transition rates as in heterogeneous diffusions~\cite{metzler2014anomalous}. For sluggish random walks in discrete lattices, the waiting times become increasingly large as the walker is displaced further from the origin.

In its simplest form (discrete time and discrete space), a sluggish random walk is a Markov chain on a one-dimensional lattice with uniform spacing $b$, where the hopping probability depends on the particle's current position $x \in \mathbb{Z}$. At each time step, the walker randomly moves one lattice spacing $b$ to the left, right, or remains at its current position, but importantly, the hopping probability decreases as the walker strays further from the origin $x_0$~\cite{del2025generalized}. The probabilities of hopping left $p_{\gets}(x)=P(X_{t+1}=x - b | X_t=x)$ and right $p_{\to}(x)=P(X_{t+1}=x + b | X_t=x)$ are time-independent but depend crucially on the walker's current position $x$ as 
\begin{equation}
    p_{\gets}(x) = p_{\to}(x) =\displaystyle \frac{1}{\left(\dfrac{|x-x_0|}{b}\right)^\alpha + 2},
    \label{eqn:hop_prob}
\end{equation}
while the probability of remaining at the current site $x$ reads $1-(p_{\gets}(x)+p_{\to}(x))$. Here, $\alpha \geq 0$ is the so-called {\em sluggishness} parameter. For $\alpha > 0$, the hopping probabilities tend to zero as $|x-x_0| \to \infty$, causing the walker to spend increasingly long times away from the origin. The hopping probabilities~\eqref{eqn:hop_prob} are centred at $x_0$, which is the initial position of the walker.

The  continuum limit of the discrete sluggish random walk yields a heterogeneous  diffusion process governed by a Fokker--Planck equation with time-independent but spatially inhomogeneous transport coefficients~\citep{zodage2023sluggish, del2025generalized, Fa_2003, Cherstvy_2013}
\begin{equation}
    \dfrac{\partial P(x, t)}{\partial t} = \dfrac{\partial^2 }{\partial x^2}\left[\dfrac{1 }{|x|^\alpha} P(x, t)\right].
    \label{eqn:slrwfokker}
\end{equation}
Equation~\eqref{eqn:slrwfokker} can equivalently be rewritten as
\begin{equation}
    \frac{\partial P(x,t)}{\partial t} = \frac{\partial}{\partial x}\!\left[D(x)\, \frac{\partial P(x,t)}{\partial x}\right] + \frac{\partial}{\partial x}\!\left[U'(x)\,P(x,t)\right],
    \label{eqn:FP_drift}
\end{equation}
with a space-dependent diffusion constant and a repulsive effective potential given respectively by~\citep{del2025generalized}
\begin{equation}
    D(x) = |x|^{-\alpha},\quad U(x) = |x|^{-\alpha}.
\end{equation}
The relation $U'(x)=D'(x)=-\alpha\,\mathrm{sgn}(x)|x|^{-\alpha-1}$ follows from $\partial_x^2(DP)=\partial_x(D\,\partial_xP)+\partial_x(D'P)$, and $U$ is an effective potential representing the spatially varying diffusivity. Here, and throughout this work, we adopt dimensionless units in which the lattice spacing $b$ and the elementary time step $\Delta t$ are set to unity, so that $D(x)=|x|^{-\alpha}$ is dimensionless; at $\alpha=0$, where the lattice walk has $D=b^2/(3\Delta t)$, $D=1$ corresponds to measuring time in units of $3\Delta t$. The physical dimensions may be restored by replacing $x\to x/b$ and $t\to t/\Delta t$. The It\^{o} stochastic differential equation associated with the Fokker--Planck equation~\eqref{eqn:slrwfokker} reads
\begin{equation}
    \dot{x}(t) = \sqrt{2D(x)}\;\eta(t)
               = \sqrt{\frac{2}{|x(t)|^\alpha}}\;\eta(t),
    \label{eqn:langslurw}
\end{equation}
where $\eta(t)$ is a zero-mean Gaussian white noise with $\langle\eta(t)\rangle=0 $ for all $t$ and autocorrelation $\langle\eta(t)\eta(t')\rangle=\delta(t-t')$. The position-dependent noise amplitude $\sqrt{2/|x|^\alpha}$ decays to zero as $|x|$ increases, progressively suppressing large displacements of the sluggish random walk, with a steeper decay the larger is the sluggishness parameter $\alpha$, whereas it diverges at the origin for $\alpha>0$.

As shown in Ref.~\cite{zodage2023sluggish}, the position probability distribution $P(x,t)$ at all times $t$ may be derived analytically by solving the Fokker--Planck equation~\eqref{eqn:slrwfokker}  subject to the initial condition $P(x,0) = \delta(x - x_0)$, for a particle starting at position $x_0$ at time $t = 0$, and the boundary condition $P(x,t) \to 0$ as $|x| \to \infty$. Solving this equation via a scaling ansatz of the form $P(x,t) = t^{-\mu} H(x/t^{\mu})$, where $\mu = 1/(\alpha+2)$ is the scaling exponent, and $H$ is an unknown scaling function,  the problem is reduced to solving a linear second-order ordinary differential equation for $H$, namely $(|\xi|^{-\alpha}H)''=-\mu\,(\xi H)'$ with $\xi=x/t^{\mu}$~\citep{Fa_2003}, which admits an exact solution. Substituting the explicit value of~$\mu$ back into the ansatz and imposing normalisation then yields the full position probability distribution at all times $t$,
\begin{equation}
    P(x,t) = \frac{\left( \frac{1}{\alpha + 2} \right)^{\frac{\alpha}{\alpha + 
    2}}}{2 \Gamma \left( \frac{\alpha + 1}{\alpha + 2} \right)}  \frac{ 
    |x|^{\alpha} }{ t^{\frac{\alpha + 1}{\alpha + 2}} }  \exp \left(
    -\frac{ |x|^{\alpha + 2} }{ (\alpha + 2)^2 t }\right), \label{eqn:svapd}
\end{equation}
where $\Gamma(x)$ is the Gamma function. For $\alpha = 0$, setting $\Gamma(1/2) = \sqrt{\pi}$, the distribution~\eqref{eqn:svapd} reduces to the Gaussian propagator $P(x,t) = (4\pi t)^{-1/2}\exp[-x^2/(4t)]$ of a simple diffusion with $D = 1$, consistent with $D(x) = |x|^{-\alpha}$ at $\alpha = 0$. For $\alpha > 0$, the distribution is starkly non-Gaussian and bimodal, characterised by two symmetric peaks located away from the origin and a cusp (vanishing density) at $x = 0$, which reflects the fact that the time spent around $x$ is proportional to $1/D(x)=|x|^{\alpha}$ and hence vanishes at the origin (see left panel in Fig.~\ref{fig:2}). The mean square displacement follows directly from~\eqref{eqn:svapd} as~\cite{zodage2023sluggish, del2025generalized}
\begin{equation}
    \langle (x (t)- x_0)^2 \rangle \propto t^{\,2/(\alpha+2)}
    = t^{\,2\mu},
    \label{eqn:MSD_sluggish}
\end{equation}
where we have introduced
\begin{equation}
    \mu = \frac{1}{\alpha+2} .
\end{equation}
Since $\mu < 1/2$, for all $\alpha > 0$,  Eq.~\eqref{eqn:MSD_sluggish} reveals the strict {\em subdiffusive} nature of sluggish random walks as the typical displacement scales with time as $\sim t^{\mu}$ i.e., slower than in Brownian diffusion $x(t)\sim t^{\frac{1}{2}}$~\citep{Einstein_1905, Lemons_1997}. Interestingly,  the degree of subdiffusion is continuously tunable through the parameter $\alpha$, making the sluggish random walk an excellent case study of the subdiffusive-to-diffusive transition with analytically solvable statistics.

\subsection{Diffusion with stochastic resetting}
\label{sec22}
\smallskip
The framework of stochastic resetting was first illustrated with a model of a Brownian diffusion on continuous space $\mathbb{R}$ with sudden events that return the diffusion to its initial point $x_0$ at random times generated by a Poisson process of rate $r$, such that inter-reset times are exponentially distributed with mean $1/r$~\citep{evans2011diffusion}. The probability density $P(x,t|x_0)$ obeys the Fokker--Planck with source terms
\begin{equation}
    \frac{\partial P(x, t|x_0)}{\partial t}
        = D\,\frac{\partial^2 P(x, t|x_0)}{\partial x^2}
          - r\,P(x, t|x_0)
          + r\,\delta(x-x_0),
    \label{forkersr}
\end{equation}
where the term $-r\,P(x,t|x_0)$ drains probability from every location at constant rate $r$, and the term $r\,\delta(x-x_0)$ re-injects probability at $x_0$ at a rate $r$. For $r > 0$, resetting breaks detailed balance and drives the system into a nonequilibrium stationary state~\citep{Gupta_2022}. Solving~\eqref{forkersr} yields a bilateral exponential stationary distribution~\citep{evans2011diffusion}
\begin{equation}
    P_{\mathrm{st}}(x|x_0)
        = \frac{1}{2}\sqrt{\frac{r}{D}}
          \exp\!\left(-\sqrt{\frac{r}{D}}\,|x-x_0|\right),
    \label{eqn:svapdr}
\end{equation}
centred at $x_0$ with characteristic length scale $\ell = \sqrt{D/r}$. See also Ref.~\cite{tal2020experimental} for an experimental realisation of diffusion with stochastic resetting with optically-trapped colloidal particles.

\subsection{Diffusion with threshold resetting}
\label{sec23}
\smallskip
Diffusions complemented with threshold resetting when the position crosses a predefined boundary~\citep{Bhat_2016} offer an alternative case study to Poissonian resetting. In this scenario, the particle starts at the origin, diffuses freely until it reaches one of the thresholds $\pm x_{\rm th}$, with $|x_{\rm th}| > x_0$, and then instantaneously resets to the origin $x_0$. Unlike Poissonian resetting, each reset cycle is a first-passage event~\cite{Redner_2001}, so the inter-reset time distribution is set entirely by the first-passage time to $x_{\rm th}$~\citep{Pal_2017}. The probability density $P(x,t)$ can be derived using renewal theory~\citep{Pal_2017, Evans_2020}, where trajectories are modelled with absorbing boundaries at $\pm x_{\rm th}$. An interesting quantity is the absorbing propagator $P_{\rm ab}(x,t)$, defined as the probability density of a freely diffusing particle starting at the origin that reaches position $x$ at time $t$ without having previously touched the boundaries $\pm x_{\rm th}$. Its Laplace transform $\tilde{P}_{\rm ab}(x,s)=\int_0^{\infty}dt e^{-st} P_{\rm ab}(x,t)$ is accessible analytically and given  by~\citep{Kantor_2007, Tuckwell_1995}
\begin{equation}
    \tilde{P}_{\rm ab}(x,s)
        = \frac{\displaystyle\sinh\!\left(\sqrt{s/D}\,
          (x_{\rm th} - |x|)\right)}
               {\sqrt{4Ds}\,
          \cosh\!\left(
          \displaystyle\sqrt{s/D}\,x_{\rm th}\right)},
    \label{eq:absorbing_prop}
\end{equation}
where $|x| < x_{\rm th}$. Equation~\eqref{eq:absorbing_prop} solves the diffusion equation with boundary conditions $\tilde{P}_{\rm ab}(\pm x_{\rm th},s) = 0$ for symmetric thresholds and initial position set to the origin $x_0 = 0$. Applying renewal theory then yields the full propagator in Laplace space~\citep{Evans_2020, Pal_2017}
\begin{equation}
    \tilde{P}(x,s) = \frac{\tilde{P}_{\rm ab}(x,s)}{s\,\tilde{P}_{\rm su}(s)},
    \label{eq:renewal_propagator}
\end{equation}
where $\tilde{P}_{\rm su}(s)$ is the Laplace transform of the survival probability, which in this scenario equals the probability that no reset has occurred up to time $t$. Taking the  limit  $\lim_{s\to 0} s\,\tilde{P}(x,s)$ gives the stationary distribution~\citep{Pinsky_2020, Bhat_2016}
\begin{equation}
    P_{\rm st}(x)
        = \begin{cases}
              \displaystyle\frac{x_{\rm th} - |x|}{x_{\rm th}^2},
                  & |x| < x_{\rm th}, \\[6pt]
              0,  & |x| \geq x_{\rm th}
          \end{cases},
    \label{eq:final_den}
\end{equation}
characterised by a triangular profile that peaks at the origin and vanishes at the boundaries.

\section{Sluggish random walk with resetting}
\label{sec:3}

Section~\ref{sec21} introduced sluggish diffusions, which slow a walker down the further it strays from the origin, and Sections~\ref{sec22} and~\ref{sec23} introduced stochastic resetting, which periodically returns a walker to a fixed point regardless of how far it has travelled. Combining sluggishness with resetting raises a natural question with no simple answer \textit{a priori}, namely, whether the resetting mechanism simply acts on top of the sluggish dynamics or whether the position-dependent mobility qualitatively reshapes the manner in which resetting drives the system to a nonequilibrium stationary state. 

In this section, we construct the two coupled models needed to answer this question. Section~\ref{sec:3.1} complements the discrete-time sluggish random walk with a Poissonian resetting probability at each time step and derives the corresponding modified Fokker--Planck equation in the continuum limit. Section~\ref{sec:3.2} instead couples the sluggish dynamics to a deterministic, threshold-based reset and rationalises why its nonlocal boundary condition requires renewal theory rather than a direct partial differential equation approach. These two models form the basis for the exact stationary distributions, first-passage statistics, and resetting costs derived in the rest of this paper.

\smallskip
\subsection{Sluggish random walk with Poissonian resetting}\label{sec:3.1}
The introduction of the resetting probability $p_r$ (Poissonian resetting) alters the dynamics of a sluggish random walk. At each discrete time step, the particle either resets to its initial position $x_0$ with probability $p_r$, or executes a step according to the sluggish random walk rules with probability $1-p_r$. Hence, the update rule of the walker's position reads
\begin{equation}
     x_{t+1} = \begin{cases}
     x_0 & \text{with probability } \; p_r\\
     x_t + b & \text{with probability } \; (1-p_r)\, p_{\to}(x_t)\\
     x_t - b & \text{with probability } \; (1-p_r)\, p_{\gets}(x_t)\\
     x_t & \text{with probability } \; (1-p_r) \Big[1- \Big(p_{\gets}(x_t) + p_{\to}(x_t)\Big)\Big]
 \end{cases},
\end{equation}
where $p_{\to}(x_t)=p_{\gets}(x_t)$ are given by Eq.~\eqref{eqn:hop_prob} evaluated at $x=x_t$, see Fig.~\ref{fig:1}A for an illustration. 

In the continuum limit, the resetting probability $p_r$ is replaced by a resetting rate $r\,dt$, so that the probability of a reset in an infinitesimal time interval $[t, t+dt]$ is $r\,dt$, recovering $r$ as a rate with units of inverse time. We propose a Fokker--Planck equation 
\begin{equation}
\frac{\partial P(x, t|x_0)}{\partial t} = \dfrac{\partial^2 }{\partial x^2}\left[\dfrac{1 }{|x|^\alpha} P(x, t)\right] - rP(x, t|x_0) + r\delta(x-x_0). \label{mainswwsr}
\end{equation}
that effectively combines the sluggish random walk with subdiffusive spread~\eqref{eqn:slrwfokker} and the well-established theory of stochastic resetting~\eqref{forkersr}. We use Eq.~\eqref{mainswwsr}, written for $x_0=0$, in the next sections as a basis for our calculations on statistics regarding fixed-time ensembles.

\subsection{Sluggish random walk with threshold resetting}\label{sec:3.2}
\begin{figure}[h]
    \centering\includegraphics[width=\textwidth]{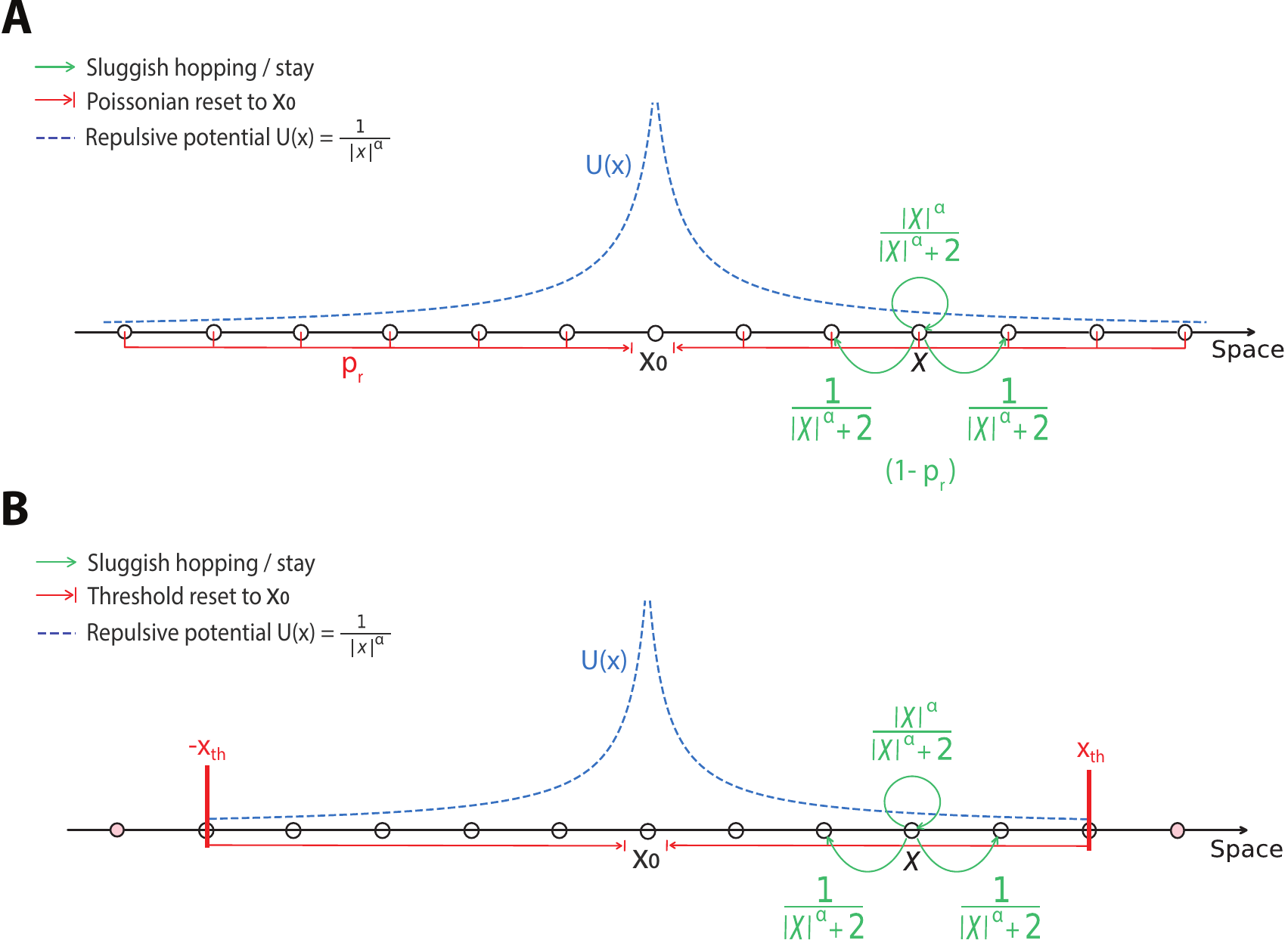}
    \caption{Schematic illustration of sluggish random walkers with resetting on a discrete one-dimensional lattice with homogeneous spacing. \textbf{(A)} Sluggish random walker with Poissonian resetting. At each position $x$, the walker resets to the origin with probability $p_r$ (red), or, with probability $1-p_r$, hops left/right with probability $1/(|x|^{\alpha}+2)$ or stays at $x$ with probability $|x|^{\alpha}/(|x|^{\alpha}+2)$. The dashed curve shows the repulsive potential $U(x)=1/|x|^{\alpha}$. \
    \textbf{(B)}  Sluggish random walker with threshold resetting. At each position $x$, the walker hops or stays with sluggish probabilities (green arrows). Upon reaching the thresholds $\pm x_{th}$ (red walls), the walker is deterministically reset to the origin $x_0$ (red arrows). In both panels, the blue dashed line sketches the effective potential $U(x)$ that follows from the continuum-limit description.} 
    \label{fig:1}
\end{figure}

\smallskip
An alternative approach developed here combines the position-dependent mobility characteristic of sluggish random walks with a deterministic threshold-based resetting mechanism~\cite{Pal_2019, del2025generalized}. A deterministic reset occurs when the walker reaches one of the thresholds, $|x(t)-x_0|\ge x_{\text{th}}$ (see Fig.~\ref{fig:1}B), after which it is placed at the reset point $x_0$ ($|x_0|<x_{\rm th}$). This resetting mechanism differs fundamentally from stochastic resetting in that it is neither random in time nor probabilistic in nature, but rather represents a state-dependent intervention that activates only when specific spatial conditions are satisfied \citep{evans2011diffusion}. This feedback mechanism leads to a confinement region $[-x_{\text{th}}, x_{\text{th}}]$ that the walker never leaves during its time evolution. For discrete-time implementations, the update rule governing the walker's dynamics is expressed as
\begin{equation}
x_{t+1} =
\begin{cases}
x_0, 
& \text{if } x_t=\pm x_{\mathrm{th}}, \\[10pt]
\begin{aligned}
x_t+b, \, & \text{ with probability } p_{\to}(x_t),\\
x_t-b, \,& \text{ with probability } p_{\gets}(x_t),\\
x_t,   \,& \text{ with probability } 1-\Big(p_{\gets}(x_t)+p_{\to}(x_t)\Big)
\end{aligned}
& \text{if } |x_t|<x_{\mathrm{th}} .
\end{cases}
\end{equation}
where $p_{\to}(x_t)=p_{\gets}(x_t)$ are given by the sluggish dynamics' position-dependent transition probabilities given by  Eq.~\eqref{eqn:hop_prob} evaluated at $x=x_t$. As we show later in Section~\ref{sec:4.2} in the continuum limit, the reset introduces a nonlocal boundary condition, whereby the probability flux that leaves the system at $\pm x_{\rm th}$ is reinjected at $x_0$~\cite{Benichou_2011}. This challenges the straightforward application of conventional partial differential equation methodologies, such as the modified Fokker--Planck equation. The inter-reset times are first-passage times to the thresholds and are therefore not exponentially distributed.

\section{Position distributions}
\label{sec:4}
\smallskip
\subsection{Position distribution of Sluggish random walk with Poissonian resetting}\label{sec:4.1}
\smallskip
In the long-time limit, the walker reaches a stationary state $P_{\text{st}}(x)$ where the probability density becomes independent of time \citep{Soize_1994}. For resetting to the origin ($x_0 = 0$), the Fokker--Planck equation~\eqref {mainswwsr} reduces to its steady-state form

\begin{equation}
\frac{d^2}{dx^2}\left[\frac{1}{|x|^{\alpha}}P_{\text{st}}(x)\right] - rP_{\text{st}}(x) + r\delta(x) = 0. \label{eq:steady_state}
\end{equation}
Since the resetting point coincides with the origin, the problem has even symmetry $P_{\text{st}}(-x) = P_{\text{st}}(x)$, so it is sufficient to solve Eq.~\eqref{eq:steady_state} on $x > 0$ and extend to the negative domain by symmetry. Away from the origin, the delta-function source vanishes and the equation becomes homogeneous. Introducing the substitution $q(x) \equiv x^{-\alpha} P_{\text{st}}(x)$, the rescaled coordinate $z = \frac{2\sqrt{r}}{\alpha + 2}\,x^{(\alpha+2)/2}$, and the further ansatz $q(x) = x^{1/2} f(z)$ (which removes the coordinate singularity at $x = 0$), transforms the homogeneous equation into the modified Bessel equation~\citep{Porter_1923} of order $\nu = 1/(\alpha+2)$,
\begin{equation}
f''(z)+\frac{1}{z}f'(z)-\left(1+\frac{\nu^2}{z^2}\right)f(z)=0.
\end{equation}
The general solution is the linear combination $f(z) = C_1 I_\nu(z) + C_2 K_\nu(z)$, where $I_\nu(z)$ and $K_\nu(z)$ denote the modified Bessel functions of the first and second kind, respectively. Normalisability of~$P_{\text{st}}(x)$ as $x \to \infty$ dictates discarding the exponentially growing branch $I_\nu$, so $C_1 = 0$, leaving $K_\nu$ as the physical solution and yielding 
\begin{equation}
    P_{\text{st}}(x) = C\,|x|^{\alpha+1/2}K_\nu\!\left(\frac{2\sqrt{r}}{\alpha+2}|x|^{(\alpha+2)/2}\right),
\end{equation}
for $x > 0$, extended to $x<0$ by the even symmetry $P_{\text{st}}(-x) = P_{\text{st}}(x)$. The normalisation constant $C$ is fixed by the jump condition that balances the diffusive flux against the resetting source at the origin. Integrating~\eqref{eq:steady_state} over $x \in (-\epsilon, \epsilon)$ and taking $\epsilon \to 0$ gives
\begin{equation}
\left.\frac{d}{dx}\!\left[|x|^{-\alpha}P_{\text{st}}(x)\right]\right|_{0^+}
- \left.\frac{d}{dx}\!\left[|x|^{-\alpha}P_{\text{st}}(x)\right]\right|_{0^-}
= -r,
\label{eq:jump_main}
\end{equation}
which, by symmetry, reduces to $2\,\frac{d}{dx}\!\left[x^{-\alpha}P_{\text{st}}(x)\right]_{0^+} = -r$. Evaluating this derivative requires the small-argument expansion of $K_\nu$ beyond its leading term. Writing $K_\nu(z)=\frac{\pi}{2\sin(\nu\pi)}\left[I_{-\nu}(z)-I_{\nu}(z)\right]$ and using $I_{\pm\nu}(z)\simeq (z/2)^{\pm\nu}/\Gamma(1\pm\nu)$, the leading behaviour $K_\nu(z)\simeq\frac{\Gamma(\nu)}{2}(z/2)^{-\nu}$ renders $x^{-\alpha}P_{\text{st}}(x)$ constant at the origin and therefore does not contribute to the derivative, whereas the term stemming from $I_\nu$ is linear in $x$ and gives
\begin{equation*}
\left.\frac{d}{dx}\!\left[x^{-\alpha}P_{\text{st}}(x)\right]\right|_{0^+}
=-C\,\frac{(\alpha+2)}{2}\,\Gamma\!\left(\frac{\alpha+1}{\alpha+2}\right)\left(\frac{\sqrt{r}}{\alpha+2}\right)^{\frac{1}{\alpha+2}}.
\end{equation*}
Solving~\eqref{eq:jump_main} for $C$ then yields the stationary distribution
\begin{align}
P_{\text{st}}(x) = \frac{r^{\frac{2\alpha+3}{2(\alpha+2)}}}{\Gamma\!\left(\frac{\alpha+1}{\alpha+2}\right)(\alpha+2)^{\frac{\alpha+1}{\alpha+2}}}
\,|x|^{\alpha+1/2}\,
K_{\frac{1}{\alpha+2}}\!\left(\frac{2\sqrt{r}}{\alpha+2}\,|x|^{(\alpha+2)/2}\right). \label{eqn:bessel}
\end{align}
The normalisation condition $\int_{-\infty}^{\infty}P_{\rm st}(x)dx=1$ (which for symmetry $P_{\text{st}}(-x) = P_{\text{st}}(x)$ reduces to $2\int_{0}^{\infty}P_{\rm st}(x)dx=1$) for the stationary distribution given by Eq.~\eqref{eqn:bessel} for any $\alpha\geq 0$ may be checked by using the changes of variable $\zeta = [2\sqrt{r}/(\alpha+2)]x^{(\alpha+2)/2}$, $\nu=1/(\alpha+2)$ and $s =2-\nu= \frac{2\alpha+3}{\alpha+2}$, and the standard Mellin-type integral of the modified Bessel function of the second kind~\citep{Erdelyi_1954, Gradshteyn_2007} $\int_0^\infty \zeta^{s-1}K_\nu(\zeta)\,d\zeta = 2^{s-2}\,\Gamma\!\left(\frac{s-\nu}{2}\right)\Gamma\!\left(\frac{s+\nu}{2}\right)$ which holds for $s > |\nu|$.

\begin{figure}
    \centering
    \includegraphics[width=1.01\textwidth, trim=0cm 0cm 0cm 0cm, clip]{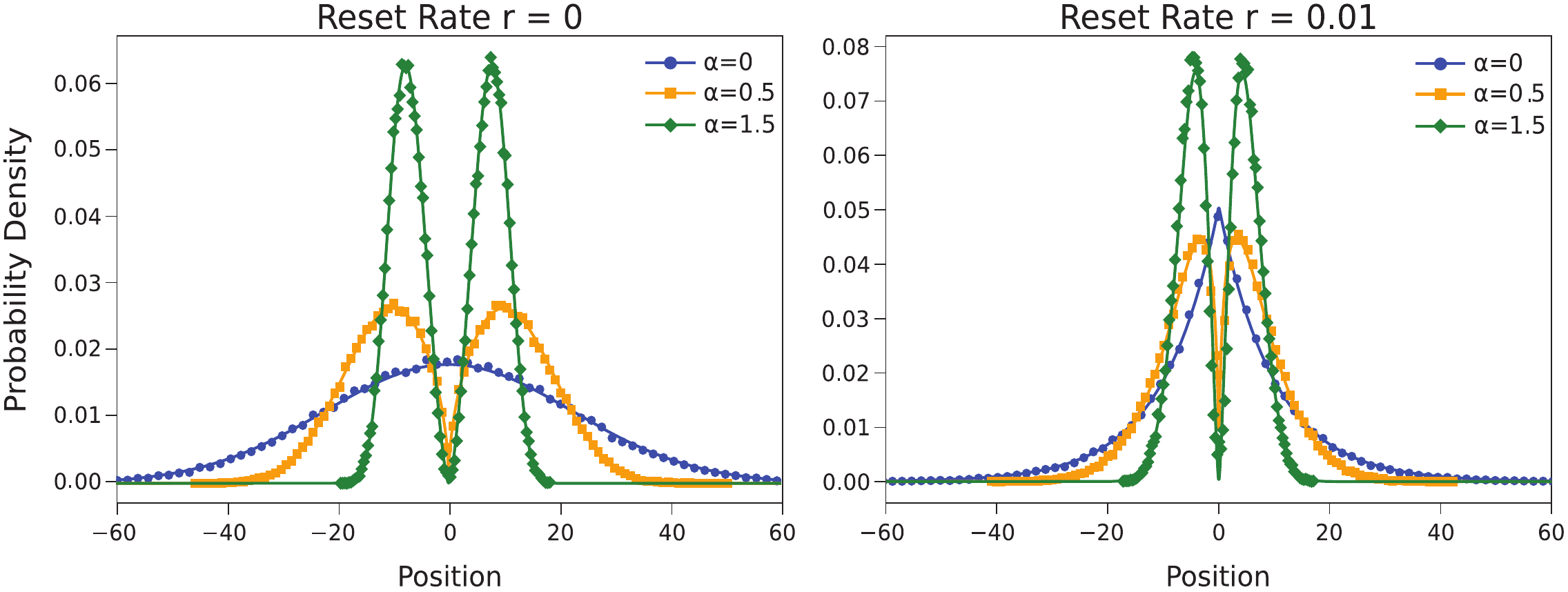}
    \caption{Steady-state probability density of sluggish random walks at the final simulation time, for reset rates $r = 0$ (left panel) and $r = 0.01$ (right panel) and sluggishness parameters $\alpha =0,\, 0.5,\,\text{and}\, 1.5$ (see legend). Symbols are obtained from numerical simulations, and lines are given by the analytical results~\eqref{eqn:bessel}. Simulations start from $x_0=0$ at $t=0$ and $N=10^6$ independent runs are executed. The observation time is $T=250$ for $r=0$ and $T=15/r=1500$ for $r=0.01$, long enough for the walkers to be stationary ($e^{-rT}\simeq3\times10^{-7}$ of the walkers have not yet reset). The case $r = 0$ retrieves exactly the result for the free sluggish random walk~\eqref{eqn:svapd}, see Ref.~\cite{zodage2023sluggish}.}
    \label{fig:2}
\end{figure}

Notably, the stationary distribution~\eqref{eqn:bessel} results from the competition between the sluggish dynamics, which makes the walker leave the high-diffusivity region around the origin and spend long times at larger distances where it moves slowly, and the resetting mechanism, which continuously reinjects probability at the origin (see Fig.~\ref{fig:2}). For $\alpha = 0$ and $r > 0$, the stationary distribution~\eqref{eqn:bessel} reduces to the bilateral exponential~\eqref{eqn:svapdr}, peaked sharply at the origin. As $\alpha$ increases at fixed $r$, the density at the resetting point (origin) is increasingly depleted, since the walker spends little time in the region of large diffusivity around the origin, and the distribution develops the same bimodal, cusped structure at $x=0$ already seen in the sluggish random walk in the absence of resetting~\eqref{eqn:svapd}. Resetting arrests the expansion of the free walk, whose spread grows as $t^{\mu}$, at the length scale $\ell=r^{-1/(\alpha+2)}$ introduced below, and the tails decay as the compressed exponential $\exp\left[-\frac{2\sqrt{r}}{\alpha+2}|x|^{(\alpha+2)/2}\right]$, which is faster than exponential for $\alpha>0$. For $r<1$ the length scale $\ell$ decreases as $\alpha$ increases, so that at fixed $r$ the distribution becomes narrower.

The Poissonian stationary distribution admits an exact scaling collapse. Introducing the rescaled variable
\begin{equation}
    z = x\,r^{1/(\alpha+2)} = \frac{x}{\ell}, \qquad \ell \equiv r^{-1/(\alpha+2)},
\end{equation}
and defining the rescaled density $P_{\text{st}}(z)$ via $P_{\rm st}(x) = r^{1/(\alpha+2)}\,P_{\text{st}}(z)$, substitution into~\eqref{eq:steady_state} eliminates all dependence on $r$, yielding the universal equation $\frac{d^2}{dz^2}[|z|^{-\alpha}P_{\text{st}}(z)] - P_{\text{st}}(z) = -\delta(z)$. Solving as before gives the closed form
\begin{equation}
    P_{\text{st}}(z) = \frac{|z|^{\alpha+1/2}}{\Gamma\!\left(\frac{\alpha+1}{\alpha+2}\right)(\alpha+2)^{\frac{\alpha+1}{\alpha+2}}}\,
    K_{\frac{1}{\alpha+2}}\!\left(\frac{2}{\alpha+2}\,|z|^{(\alpha+2)/2}\right),
    \label{eqn:bessel_scaled}
\end{equation}
independent of $r$. This scaled distribution evolves continuously with $\alpha$, developing sharper bimodal peaks that migrate further from the origin as sluggishness increases (see Fig.~\ref{fig:3}A).

\subsection{Position distribution of Sluggish random walk with threshold resetting}\label{sec:4.2}
\begin{figure}[h]
    \centering
    \includegraphics[width=1\textwidth, trim=0cm 0cm 0cm 0cm, clip]{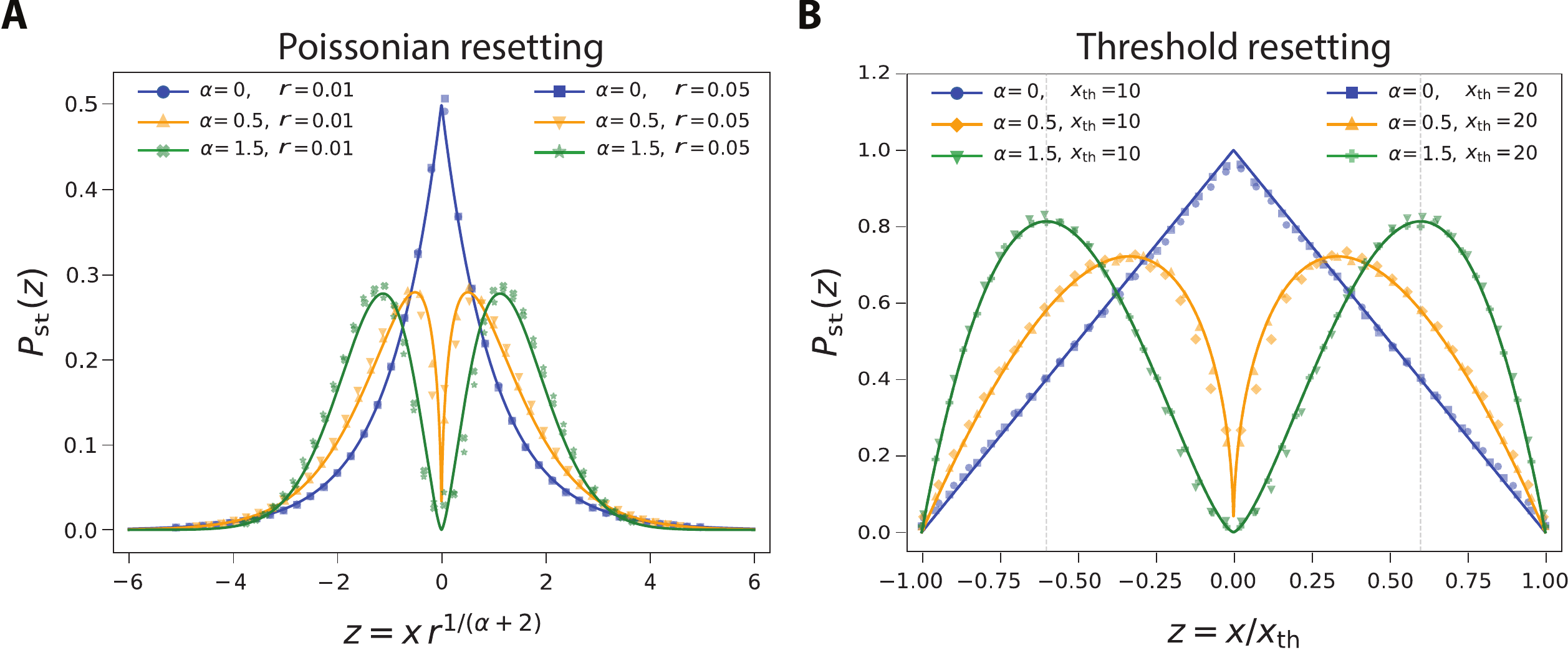}
    \caption{Scaled steady-state probability density of sluggish random walks with resetting. Symbols are obtained from numerical simulations, and lines are given by the corresponding analytical result. Simulations start from $x_0=0$ and $N=10^6$ independent realisations are used. Final positions are taken at $T=15/r$ in (A) and at $T=10\langle\tau\rangle$ in (B), with $\langle\tau\rangle$ given by Eq.~\eqref{eq:cycle_time}, so that the density is stationary. \textbf{(A)} Poissonian resetting, $P_{\text{st}}(z)$~\eqref{eqn:bessel_scaled}, plotted as a function of the rescaled position $z = x\,r^{1/(\alpha+2)}$, for sluggishness parameters $\alpha \in \{0,\, 0.5,\, 1.5\}$ and reset rates $r \in \{0.01,\, 0.05\}$ (see legend). The collapse of simulations at different reset rates $r$ onto a single curve for each $\alpha$ confirms the $r$-independence of $P_{\text{st}}(z)$. \textbf{(B)} Threshold resetting, $P_\mathrm{st}(z)$~\eqref{eqn:posfeedscale}, plotted as a function of the rescaled position $z = x/x_\mathrm{th}$, for sluggishness parameters $\alpha \in \{0,\, 0.5,\, 1.5\}$ and thresholds $x_\mathrm{th} \in \{10,\, 20\}$ (see legend). The vertical dashed lines mark the universal peak location $|z^*| = 0.6$ for $\alpha = 1.5$.}
    \label{fig:3}
\end{figure}

\smallskip
The deterministic, threshold-based nature of the resetting mechanism implies that the steady-state distribution can be obtained using a renewal-theory framework~\citep{Singh_2020}. The central idea is that the total probability density decomposes into contributions from trajectories that have never reached the threshold and those that have undergone at least one reset event, yielding the renewal equation
\begin{equation}
P(x,t) = P_0(x,t) + \int_0^t \rho_{\rm th}(t')\, P(x,t-t')\, dt',
\end{equation}
where $P_0(x,t)$ is the absorbing propagator of walkers that started at the origin and have not yet reached either threshold, and $\rho_{\rm th}(t')$ is the probability density that the walker, starting from the origin, first reaches the threshold $|x| = x_{\text{th}}$ at time $t'$ \citep{Roldan_2017}, in other words, the first-passage-time distribution to reach any of the thresholds $\pm x_{\rm th}$. Taking the Laplace transform and applying the final value theorem, the stationary distribution reads
\begin{equation*}
P_{\text{st}}(x) = \lim_{s \to 0} \frac{\tilde{P}_0(x,s)}{\tilde{Q}_{\rm th}(s)},
\end{equation*}
where $\tilde{Q}_{\rm th}(s)$ is the Laplace transform of the survival probability against reaching the threshold. This reduces the problem to computing $u(x) = \int_0^\infty P_0(x,t)\,dt$, the Green's function of the absorbed sluggish process representing the expected time the walker spends near $x$ before first reaching the threshold, which satisfies
\begin{equation}
  \frac{d^2}{dx^2}\left[|x|^{-\alpha}\, u(x)\right] = -\delta(x), \quad u(\pm x_{\text{th}}) = 0,\label{eq:23}
\end{equation}
with absorbing boundary (Dirichlet) conditions at the thresholds. The substitution $v(x) = |x|^{-\alpha} u(x)$ reduces the problem~\eqref{eq:23} to $v''(x) = -\delta(x)$, which we solve by integrating across the singularity at the origin and imposing the absorbing boundary condition $v(\pm x_{\text{th}}) = 0$. Exploiting the even symmetry of the problem yields $v(x) = \tfrac{1}{2}(x_{\text{th}} - |x|)$, and back-substituting $v\to u$ gives $u(x) = \tfrac{1}{2}|x|^\alpha(x_{\text{th}} - |x|)$. Let $\tau$ denote the duration of a reset cycle whose probability density is $\rho_{\rm th}$. Since $\tilde{Q}_{\rm th}(0)=\int_0^\infty Q_{\rm th}(t)\,dt=\langle\tau\rangle$, the normalisation is provided by the mean cycle duration
\begin{equation}
    \langle\tau\rangle=\int_{-x_{\rm th}}^{x_{\rm th}}u(x)\,dx=\frac{x_{\rm th}^{\alpha+2}}{(\alpha+1)(\alpha+2)},
    \label{eq:cycle_time}
\end{equation}
and $P_{\rm st}(x)=u(x)/\langle\tau\rangle$ yields the closed-form stationary distribution
\begin{equation}
P_{\text{st}}(x) = 
\begin{cases} 
\dfrac{(\alpha+1)(\alpha+2)}{2} \dfrac{|x|^\alpha (x_{\text{th}} - |x|)}{x_{\text{th}}^{\alpha+2}} & |x| < x_{\text{th}}, \\[6pt]
0 & |x| \geq x_{\text{th}}.
\end{cases}
\label{eqn:posfeed}
\end{equation}
This distribution has a characteristic size set by the threshold $x_\text{th}$. By introducing the normalised variable 
\begin{equation}
    z = \frac{x}{x_\text{th}},
\end{equation} and applying the standard change-of-variables formula, the scaled distribution is given by
\begin{equation}
  P_{\text{st}}(z) = \dfrac{(\alpha+1)(\alpha+2)}{2} |z|^{\alpha} (1 - |z|), \quad \quad \text{for } |z| < 1, \label{eqn:posfeedscale}
\end{equation}
which is a special case of a symmetric Beta-type distribution on $[-1, 1]$ with parameters $(\alpha+1,\, 2)$ (see Fig.~\ref{fig:3}B). Optimising $P_{\text{st}}(z)$ with respect to $|z|$ shows that its maximum occurs at $|z|^{*} = \alpha/(\alpha+1)$, so the mode of the unscaled position is $x^{*} = \pm[\alpha/(\alpha+1)] x_\text{th}$, which moves towards $\pm x_\text{th}$ as $\alpha$ increases, meaning that large sluggishness concentrates the probability mass near the boundary rather than near the origin as it occurs in diffusions with stochastic resetting.

The scaled stationary distribution~\eqref{eqn:posfeedscale} reflects a balance set purely by the geometry of the confining interval, as shown in Fig.~\ref{fig:3}B where the shape of $P_{\text{st}}(z)$ evolves continuously with $\alpha$ across the fixed interval $|z|<1$. For $\alpha = 0$, the distribution reduces to a linear profile peaked at the origin and vanishing at the boundary, reflecting the fact that a non-sluggish walker restarts from the origin after each reset and is therefore found less often as it approaches the threshold. As $\alpha$ increases, the growing suppression of mobility away from the origin makes the walker linger disproportionately at larger $|x|$ once it leaves the origin region, and the distribution's maximum migrates from the origin towards the threshold according to $|z^*| = \alpha/(\alpha+1)$, reaching $|z^*| = 0.6$ for $\alpha = 1.5$. For threshold resetting, because the walker is only reset once it reaches $\pm x_{\text{th}}$, rather than being repeatedly pulled back to $x_0$ as in Poissonian resetting, sluggish excursions are free to accumulate probability at large $|z|$ near the interval's boundaries, pushing probability mass towards the threshold boundaries $\pm x_{\rm th}$ as $\alpha$ grows.

\section{First-passage times}\label{sec:5}
\smallskip
\subsection{First-passage time distribution}
\smallskip
We now discuss the first-passage time (FPT) defined as the earliest time at which the position reaches a prescribed target~\citep{Redner_2001}. Fixing the  initial position  to $x_0$, the FPT that we consider in this work is defined as
\begin{equation}
    t_f = \min\bigl\{t > 0 : x(t) = x_{\mathrm{tar}}\bigr\},
    \label{eq:FPT_def}
\end{equation}
and its probability density $F(t) = -dQ/dt$, where $Q(x_0,t)$ is the survival probability that the particle has not yet reached $x_{\mathrm{tar}}$ by time $t$, encodes the full statistics of search times~\citep{Redner_2001, Evans_2020}. For a freely diffusing Brownian particle in an unbounded domain, $F(t)$ decays as the power law $t^{-3/2}$ for large $t$, a heavy tail that implies a divergent mean first-passage time (MFPT)~\citep{Redner_2001}. Stochastic resetting fundamentally changes this picture by periodically returning the particle to $x_0$, which prevents arbitrarily long excursions away from the target and converts the power-law tail into an exponential decay, rendering the MFPT finite for any $r > 0$~\citep{evans2011diffusion, Evans_2020}.

For the sluggish random walk, the origin is the natural and physically distinguished target. It is the unique fixed point of the dynamics where the position-dependent diffusion coefficient $D(x) = |x|^{-\alpha}$ diverges for $\alpha>0$ and around which the stationary distribution under resetting to the origin is centred. We therefore consider a particle starting at $x_0 > 0$, with diffusivity $D(x)$ and an absorbing boundary at the origin, and returns to $x_0$ upon resetting. In the absence of resetting, the survival probability $Q_0(x_0,t)$ is the probability that the origin has not been reached by time $t$. Being a backward equation, $x_0$ plays the role of the running (differentiated) variable, and $Q_0(x_0,t)$ satisfies the backward Fokker--Planck equation~\citep{zodage2023sluggish, del2025generalized}
\begin{equation}
    \partial_t Q_0(x_0,t) = |x_0|^{-\alpha}\,\partial_{x_0}^2 Q_0(x_0,t),
    \qquad Q_0(0,t) = 0,
    \quad Q_0(x_0, 0) = 1.
    \label{eq:backward_Q0}
\end{equation}
Since the sluggish process obeys the scaling invariance $x_0 \sim t^\mu$ with $\mu = 1/(\alpha+2)$ (see Sec.~\ref{sec21}), the survival probability admits the exact scaling solution~\citep{del2025generalized}
\begin{equation}
    Q_0(x_0, t)
        = 1 - \frac{            \Gamma\!\left(\mu,\;\mu^2\,x_0^{1/\mu}/t\right)}{\Gamma(\mu)},
    \label{eq:Q0_exact}
\end{equation}
where $\Gamma(a, z) = \int_z^\infty u^{a-1} e^{-u}\, du$ is the upper incomplete gamma function~\citep{AbramowitzStegun}. As $t \to \infty$, the argument $\mu^2 x_0^{1/\mu}/t \to 0$, and since $\lim_{z\to 0}\Gamma(\mu,z) = \Gamma(\mu)$, we have $Q_0 \to 0$, confirming the recurrence of the sluggish walk in one dimension~\citep{zodage2023sluggish}. In the scaling limit $x_0 \sim \mathcal{O}(1)$, for $t$ large ($t \to \infty$), the small-argument asymptotics of the incomplete gamma function yield a power-law decay~\citep{zodage2023sluggish, del2025generalized}
\begin{equation}
    Q_0(x_0, t) \sim \frac{\mu^{2\mu-1}}{\Gamma(\mu)}\,\frac{x_0}{t^\mu},
    \qquad
    F_0(x_0,t) \equiv -\frac{\partial Q_0}{\partial t}
              \sim \frac{\mu^{2\mu}}{\Gamma(\mu)}\,\frac{x_0}{t^{1+\mu}},
    \label{eq:F0_tail}
\end{equation}
so the FPT density inherits the power-law tail $F_0 \sim t^{-1-\mu}$. The persistence exponent $\theta$, which governs the algebraic decay of the survival probability $Q_0 \sim t^{-\theta}$ is given as $\theta = \mu = 1/(\alpha+2)$ by~\eqref{eq:F0_tail} and it is consistent with the general relation between the subdiffusive scaling exponent and the persistence exponent established in~\citep{Bray_2013}; for $\alpha = 0$ one recovers the classical result $F_0 \sim t^{-3/2}$ of simple diffusion. Since $\mu \leq 1/2 < 1$ for all $\alpha \geq 0$, the mean first-passage time $\langle t_f \rangle = \int_0^\infty t\, F_0(t)\, dt$ diverges for all values of the sluggishness parameter~\citep{Redner_2001}, making resetting essential to obtain a finite search time.

\subsubsection{First passage of sluggish random walks with Poissonian resetting}
When the particle resets to its initial position $x_0 > 0$ at a constant rate $r > 0$, the dynamics between successive resets are that of the free sluggish walk with an absorbing target at the origin.  Crucially, the particle both \emph{starts} and \emph{resets} at the same point $x_0$, so every inter-reset excursion is statistically identical and the process possesses a clean renewal structure~\citep{Evans_2020}. Decomposing the trajectories according to the time $t'$ of the first reset, the survival probability $Q_r(x_0,t)$ satisfies the renewal equation~\citep{evans2011diffusion, Evans_2020}
\begin{equation}
    Q_r(x_0, t)
    = e^{-rt}\,Q_0(x_0,t)
            + r\int_0^t dt'\;e^{-rt'}\,Q_0(x_0,t')\,
        Q_r(x_0,t-t'),
    \label{eq:renewal_Qr}
\end{equation}
where the first term accounts for trajectories that experience no reset up to time~$t$, and the integral sums over the time $t'$ of the first reset, which occurs with probability density $re^{-rt'}$. Up to $t'$ the walker follows the free dynamics and avoids the target with probability $Q_0(x_0,t')$. At $t'$ it returns to $x_0$ and the process restarts, so it avoids the target during the remaining time $t-t'$ with probability $Q_r(x_0,t-t')$~\citep{Evans_2020}. Taking the Laplace transform $\widetilde{Q}(s)=\int_0^\infty e^{-st}Q(t)\,dt$ of~\eqref{eq:renewal_Qr} and solving algebraically yields
\begin{equation}
    \widetilde{Q}_r(x_0,s)
        = \frac{\widetilde{Q}_0(x_0,\,s+r)}
               {1 - r\,\widetilde{Q}_0(x_0,\,s+r)},
    \label{eq:Qr_laplace}
\end{equation}
from which all first-passage-time statistics may be retrieved.  The Laplace transform  $\widetilde{Q}_0(x_0,s)$ of the free survival probability is obtained from the exact solution~\eqref{eq:Q0_exact} via
\begin{equation}
    \widetilde{Q}_0(x_0,s)
        = \int_0^\infty e^{-st}\,Q_0(x_0,t)\,dt.
    \label{eq:Q0_laplace}
\end{equation}
This transform can be obtained in closed form. Writing $\widetilde{Q}_0=(1-\widetilde{F}_0)/s$, where $\widetilde{F}_0(x_0,s)$ is the Laplace transform of the free FPT density, Eq.~\eqref{eq:backward_Q0} becomes $s\widetilde{F}_0=x_0^{-\alpha}\partial_{x_0}^2\widetilde{F}_0$, with $\widetilde{F}_0(0,s)=1$ and $\widetilde{F}_0$ bounded as $x_0\to\infty$. Its solution gives
\begin{equation}
\begin{aligned}
    \widetilde{Q}_0(x_0,s) &= \frac{1-B(y)}{s},
    \qquad y=\frac{2\sqrt{s}\,x_0^{(\alpha+2)/2}}{\alpha+2},\\
    B(y) &= \frac{2}{\Gamma(\mu)}\left(\frac{y}{2}\right)^{\mu}K_\mu(y),
\end{aligned}
\label{eq:Q0_closed}
\end{equation}
where $K_\mu$ is the modified Bessel function of the second kind~\citep{AbramowitzStegun} and $B(y)=\widetilde{F}_0(x_0,s)$ satisfies $B(0)=1$ and decays exponentially at large $y$. For $\alpha=0$ one has $\mu=1/2$ and $K_{1/2}(y)=\sqrt{\pi/(2y)}\,e^{-y}$, so that $B(y)=e^{-y}$ and one recovers the classical result $\widetilde{Q}_0(x_0,s) = (1 - e^{-x_0\sqrt{s}})/s$ for simple diffusion with $D = 1$~\citep{evans2011diffusion}.

\begin{figure}
    \centering
    \includegraphics[width=0.85\textwidth, trim=0cm 0cm 0cm 0cm, clip]{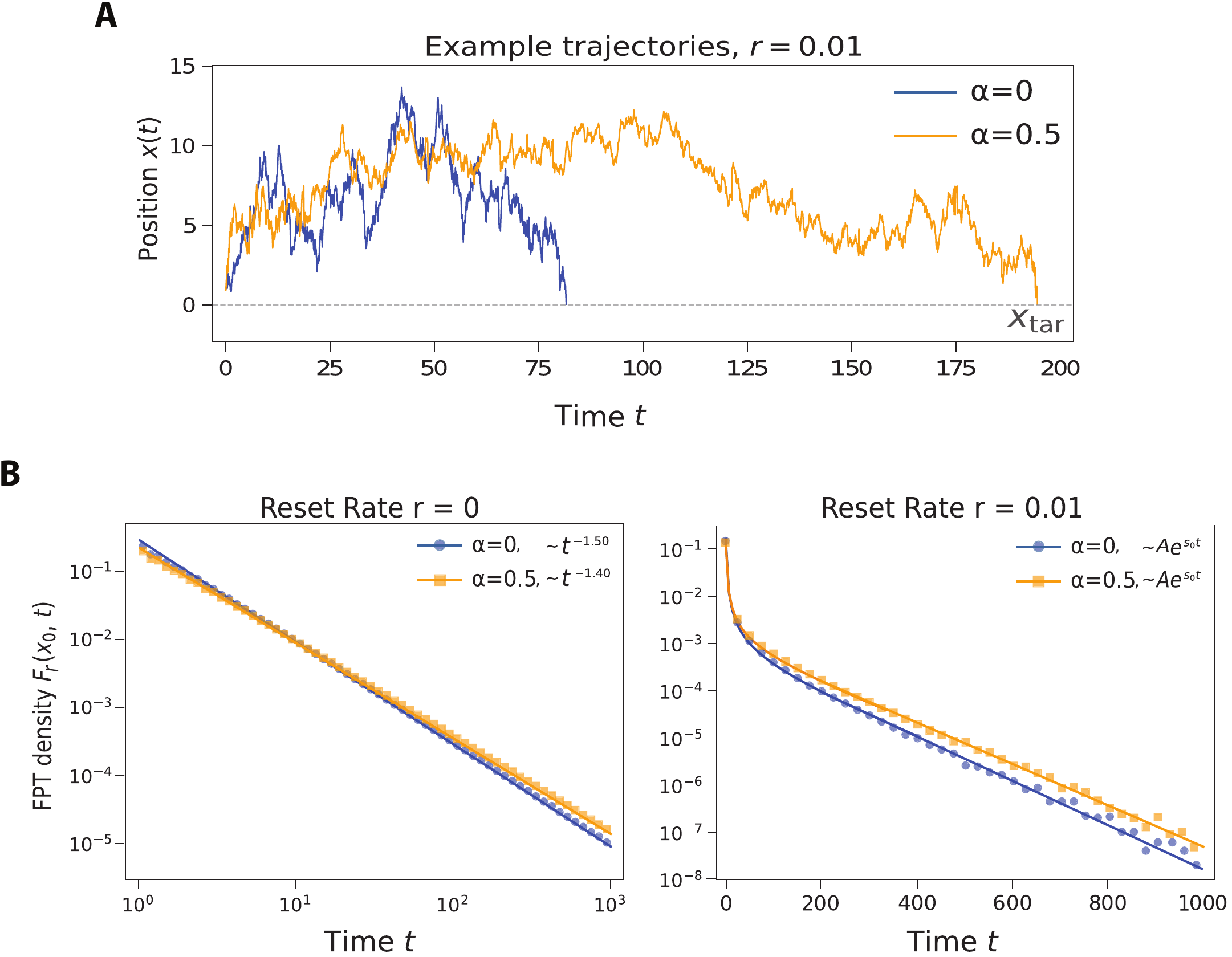}
    \caption{\textbf{(A)} Example trajectories $x(t)$ starting at $x_0 = 1$ under Poissonian resetting with $r = 0.01$, for $\alpha =0$ (random walk with resetting) and $\alpha= 0.5$ (sluggish walk with resetting), illustrating the first-passage event to the target $x_{\mathrm{tar}} = 0$ (dashed line). \ \textbf{(B)} First-passage time density $F_r(x_0,t)$ of sluggish random walks starting at $x_0 = 1$ with an absorbing target at the origin, for reset rates $r = 0$ (left panel) and $r = 0.01$ (right panel) and sluggishness parameters $\alpha \in \{0,\, 0.5\}$ (see legend). Symbols are obtained from $N = 5\times 10^6$ independent simulations evolved to $T = 1000$. For $r=0$, the lines show the analytical power-law tail $F_0 \sim t^{-(1+\mu)}$~\eqref{eq:F0_tail}. Lines are the exact first-passage densities, obtained by numerical inversion of the Laplace transform~\eqref{eq:Q0_closed} for $r=0$ and~\eqref{eq:Qr_laplace} for $r=0.01$. The densities decay as the power law $t^{-(1+\mu)}$ for $r=0$ and, at long times, exponentially for $r=0.01$.}
    \label{fig:4}
\end{figure}

At the level of the full FPT probability density, the long-time behaviour changes qualitatively from power-law to exponential decay.  The dominant singularity of $\widetilde{F}_r(s)=1-s\,\widetilde{Q}_r(s)$ in the complex $s$-plane is a simple pole at $s=s_0<0$, determined by the implicit equation~\citep{Evans_2020}
\begin{equation}
    1 - r\,\widetilde{Q}_0(x_0,\,s_0+r) = 0,
    \label{eq:pole}
\end{equation}
so that at large times $t\to\infty$ one has~\citep{Evans_2020}
\begin{equation}
    F_r(x_0,t) \sim A\,e^{s_0 t},
    \label{eq:Fr_tail}
\end{equation}
where $A>0$ is the residue of $\widetilde{F}_r$ at the pole $s_0<0$. The characteristic decay time $\tau_r=-1/s_0$ depends on both $r$ and $\alpha$ through the sluggishness-dependent Laplace transform $\widetilde{Q}_0(x_0,s)$. Using Eq.~\eqref{eq:Q0_closed}, the pole condition~\eqref{eq:pole} takes the explicit form $B\bigl(y(s_0+r)\bigr)=-s_0/r$, which has a solution in the interval $-r<s_0<0$ that is readily found numerically. For $\alpha = 0$, the pole equation admits the equivalent reformulation~\citep{evans2011diffusion} whereby, writing $s_0 = r(v_0-1)$, the quantity $v_0$ satisfies the transcendental equation
\begin{equation}
    v_0 = 1 - e^{-v_0^{1/2}\,q},
    \qquad q = x_0\sqrt{r/D},
    \label{eq:v0}
\end{equation}
which has a unique solution $0 < v_0 < 1$, and the amplitude follows analytically from the residue of $\widetilde{Q}_r$ at the pole~\citep{evans2011diffusion}
\begin{equation}
    A
        = |s_0|\frac{2v_0^{3/2}}{q\, v_0 + 2v_0^{1/2} - q},
    \label{eq:A_alpha0}
\end{equation}
which is the exact amplitude of the asymptotic tail for $\alpha=0$.  For $\alpha > 0$, the amplitude is determined from the residue formula
\begin{equation}
    A = \frac{s_0}{r^2\,\partial_s\widetilde{Q}_0(x_0,s)
        \big|_{s\,=\,s_0+r}},
    \label{eq:A_general}
\end{equation}
which can be evaluated explicitly from Eq.~\eqref{eq:Q0_closed} using $B'(y)=-\bigl(2^{1-\mu}/\Gamma(\mu)\bigr)\,y^{\mu}K_{1-\mu}(y)$. The pole term controls the exponential decay of the FPT density at long times (Fig.~\ref{fig:4}B, right panel). The branch point of $\widetilde{Q}_0(x_0,s+r)$ at $s=-r$ lies close to $s_0$ when $rx_0^{\alpha+2}\ll1$, so the branch-cut contribution, which decays as $e^{-rt}$ times a power law, becomes negligible compared with the pole term only at long times.

The mean first-passage time (MFPT) is obtained from the survival probability via integration by parts, $\langle t_f\rangle_r = \int_0^\infty Q_r(t)\,dt = \widetilde{Q}_r(x_0,0)$~\citep{Evans_2020, Gupta_2022}, giving
\begin{equation}
    \langle t_f\rangle_r
        = \frac{\widetilde{Q}_0(x_0,r)}
               {1 - r\,\widetilde{Q}_0(x_0,r)}
        = \frac{1}{r}\left[\frac{1}{B(y_r)}-1\right],
    \qquad y_r=\frac{2\sqrt{r}\,x_0^{(\alpha+2)/2}}{\alpha+2},
    \label{eq:MFPT_r}
\end{equation}
where we used $1-r\widetilde{Q}_0(x_0,r)=B(y_r)$ from Eq.~\eqref{eq:Q0_closed}. The MFPT $\langle t_f\rangle_r$ is well-defined and finite for any $r>0$, since the denominator in~\eqref{eq:MFPT_r}, which equals $B(y_r)$, is strictly positive. As $r\to0$, the small-argument expansion of $K_\mu$ gives $1-B(y)\simeq[\Gamma(1-\mu)/\Gamma(1+\mu)](y/2)^{2\mu}$, so that $\langle t_f\rangle_r\simeq\mu^{2\mu}[\Gamma(1-\mu)/\Gamma(1+\mu)]\,x_0\,r^{-(\alpha+1)/(\alpha+2)}$. This divergence reflects the infinite MFPT of the free walk, which holds for all $\alpha\geq 0$ (for $\alpha=0$ it reduces to $x_0/\sqrt{r}$). For $r\to\infty$ the MFPT also diverges, since $B(y_r)$ decays exponentially and the particle is reset before it can reach the origin. Being continuous and divergent at both ends, $\langle t_f\rangle_r$ has a minimum at an optimal resetting rate $r^*$, determined by $d\langle t_f\rangle_r/dr=0$~\citep{evans2011diffusion, Reuveni_2016}, which gives
\begin{equation}
    \frac{d}{dr}\!\left[\frac{\widetilde{Q}_0(x_0,r)}
                              {1-r\,\widetilde{Q}_0(x_0,r)}\right]
        = 0.
    \label{eq:rstar}
\end{equation}
Using Eq.~\eqref{eq:MFPT_r} and the scaled variable $y^*=2\sqrt{r^*}\,x_0^{(\alpha+2)/2}/(\alpha+2)$, the condition~\eqref{eq:rstar} can be written as
\begin{equation}
    -y^*B'(y^*)=2B(y^*)\bigl[1-B(y^*)\bigr],
    \label{eq:ystar}
\end{equation}
which contains neither $x_0$ nor $r$ separately. For $\alpha=0$, where $B(y)=e^{-y}$, Eq.~\eqref{eq:ystar} reduces to the transcendental equation~\citep{evans2011diffusion}
\begin{equation}
    \frac{q^*}{2} = 1 - e^{-q^*},
    \qquad q^* = x_0\sqrt{r^*/D},
    \label{eq:qstar}
\end{equation}
which has the unique non-zero solution $q^* \approx 1.5936$, giving the optimal rate $r^* = (q^*)^2 D/x_0^2$ for standard diffusion~\citep{evans2011diffusion}. For $\alpha > 0$, no closed-form analogue of Eq.~\eqref{eq:qstar} exists and $r^*$ must instead be obtained numerically. Since $y^*$ does not depend on $x_0$, the optimal rate obeys the scaling $r^*x_0^{\alpha+2}=[(\alpha+2)y^*/2]^2$, with $y^*$ a function of $\alpha$ only. Inserting this into Eq.~\eqref{eq:MFPT_r}, the minimal MFPT is
\begin{equation}
    \langle t_f\rangle_{r^*}=\mathcal{K}(\alpha)\,x_0^{\alpha+2},
    \qquad
    \mathcal{K}(\alpha)=\frac{4}{(\alpha+2)^2y^{*2}}\left[\frac{1}{B(y^*)}-1\right],
    \label{eq:Kalpha}
\end{equation}
i.e, $\mathcal{K}(\alpha)$ is the minimal MFPT in units of the diffusive time $x_0^2/D(x_0)=x_0^{\alpha+2}$ at the starting point. For $\alpha=0$ it reduces to $\mathcal{K}(0)=(e^{q^*}-1)/q^{*2}\approx1.54$ in units where $D=1$.

When resetting is present, the first-passage time density inherits a qualitative change in its long-time behaviour, as shown in Fig.~\ref{fig:4}B where the FPT density power-law tail [Eq.~\eqref{eq:F0_tail}] for free sluggish random walks with $r=0$ is substituted by a resetting-induced exponential decay [Eq.~\eqref{eq:Fr_tail}] for $r>0$. In the absence of resetting, changing $\alpha$ changes the power-law \emph{exponent} of the tail, since a larger $\alpha$ lowers the persistence exponent $\mu=1/(\alpha+2)$ and makes the walker take systematically longer times to first return to the origin. Once resetting is introduced at $r=0.01$, the tail instead becomes exponential, but this decay is not universal. The decay rate $s_0$ remains $\alpha$-dependent, since $s_0$ is obtained from the same $\alpha$-dependent Laplace transform $\widetilde{Q}_0(x_0,s)$ that generated the free-walk tail, so the more sluggish walker ($\alpha=0.5$) still decays slightly more slowly than $\alpha=0$ (see Fig.~\ref{fig:4}B). Example trajectories for $r=0.01$ are shown in Fig.~\ref{fig:4}A, which illustrate the first-passage event to the target. Since the time between successive resets is exponentially distributed with mean $1/r$ irrespective of $\alpha$, any difference between the two walkers arises from their spatial dynamics between resets.

\subsubsection{First-passage distribution of sluggish random walk with threshold resetting}\label{sec:5.1.2}
For the threshold resetting case, the FPT problem involves two competing absorbing boundaries.  Starting from $x_0 \in (0, x_{\rm th})$, the particle is absorbed at the origin, which constitutes a successful first passage, or it hits $x_{\rm th}$ and is reset to $x_0$, restarting the cycle.  The probability that a single excursion ends in absorption at the origin rather than at the threshold is the splitting probability $\pi(x_0)$.  Since the sluggish walk is governed by the Ito stochastic differential equation (SDE)~\citep{del2025generalized}
\begin{equation}
    dx = \sqrt{2\,|x|^{-\alpha}}\;dW,
    \label{eq:sde_sluggish}
\end{equation}
which has no deterministic drift in the Ito sense, the backward generator is simply $|x_0|^{-\alpha}\partial_{x_0}^2$~\citep{Redner_2001}. 

\noindent
The splitting probability therefore satisfies
\begin{equation}
    |x_0|^{-\alpha}\,\pi''(x_0) = 0,
    \qquad \pi(0) = 1,
    \quad \pi(x_{\rm th}) = 0,
    \label{eq:splitting_ode}
\end{equation}
which, since $|x_0|^{-\alpha} \neq 0$ for $x_0 > 0$, reduces to $\pi''(x_0) = 0$.  The general solution is linear, $\pi(x_0) = A + B x_0$, and imposing the boundary conditions gives
\begin{equation}
    \pi(x_0)
        = 1 - \frac{x_0}{x_{\rm th}},
    \qquad 0 \leq x_0 \leq x_{\rm th}.
    \label{eq:splitting}
\end{equation}
This result is independent of the sluggishness parameter $\alpha$ and coincides with the splitting probability of Brownian diffusion~\citep{Redner_2001}.  The absence of $\alpha$-dependence reflects the fact that the Ito SDE~\eqref{eq:sde_sluggish} carries no deterministic drift; $x(t)$ is a martingale and the position-dependent diffusivity $|x|^{-\alpha}$ affects the \emph{time} taken to reach a boundary but not which boundary is reached first. The two boundaries $0$ and $x_{\rm th}$ sit at unequal distances from the starting point $x_0$ unless $x_0=x_{\rm th}/2$, so $\pi(x_0)$ is generally asymmetric between them, exactly as~\eqref{eq:splitting} shows. Analogous considerations may follow from the martingale properties of the Ito SDE~\eqref{eq:sde_sluggish}, see Ref.~\cite{roldanmartingales}.

\begin{figure}
    \centering
    \includegraphics[width=1\textwidth, trim=0cm 0cm 0cm 0cm, clip]{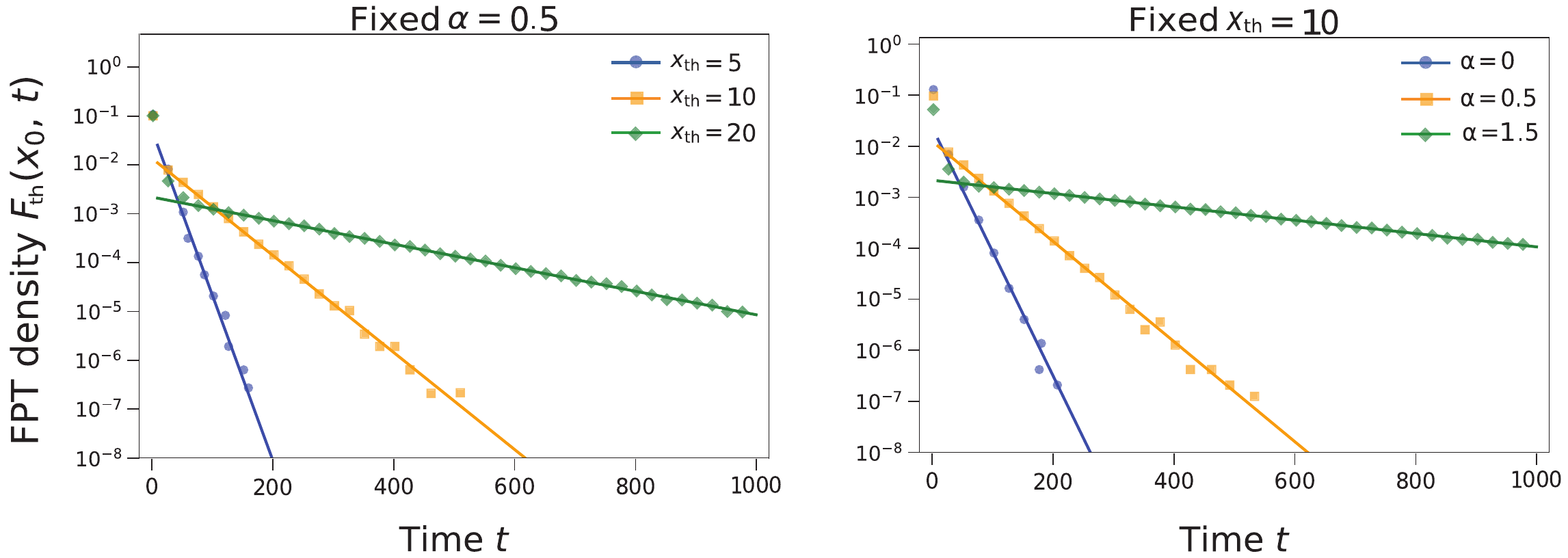}
    \caption{First-passage time density $F_{\rm th}(x_0,t)$ of sluggish random walks with threshold resetting, starting at $x_0 = 1$ with an absorbing target at the origin, for fixed $\alpha = 0.5$ with thresholds $x_{\rm th} \in \{5,\, 10,\, 20\}$ (left panel) and fixed $x_{\rm th} = 10$ with sluggishness parameters $\alpha \in \{0,\, 0.5,\, 1.5\}$ (right panel). Symbols are obtained from $N = 10^6$ independent simulations evolved to $T = 1000$, with every fifth data point in the histograms shown here. Solid lines are the exact densities (numerical inversion of~\eqref{eq:Fth_laplace}) and dashed lines the exponential tail $A_{\rm th}e^{s_{\rm th}t}$~\eqref{eq:Fth_tail}, with $s_{\rm th}$ and $A_{\rm th}$ given analytically, confirming the finite MFPT guaranteed by Eq.~\eqref{eq:MFPT_th}. Smaller thresholds and smaller sluggishness parameters both produce faster exponential decays.}
    \label{fig:5}
\end{figure}

Since the particle is confined to the finite interval $(0, x_{\rm th})$ and each excursion terminates at one of the two boundaries with probability one, the geometric renewal structure gives the MFPT to the origin as~\citep{Redner_2001}
\begin{equation}
    \langle t_f \rangle_{\rm th}
        = \frac{\langle \tau_{\rm ex}\rangle}{\pi(x_0)}
        = \frac{\langle \tau_{\rm ex}\rangle}{1 - x_0/x_{\rm th}},
    \label{eq:MFPT_th}
\end{equation}
where $\langle \tau_{\rm ex}\rangle$ is the mean duration of a single excursion. It satisfies the backward equation $x_0^{-\alpha}\partial_{x_0}^2\langle\tau_{\rm ex}\rangle=-1$, with $\langle\tau_{\rm ex}\rangle=0$ at $x_0=0$ and $x_0=x_{\rm th}$, which gives
\begin{equation}
    \langle\tau_{\rm ex}\rangle
        = \frac{x_0\bigl(x_{\rm th}^{\alpha+1}-x_0^{\alpha+1}\bigr)}{(\alpha+1)(\alpha+2)}, 
    \label{eq:tau_exit}
\end{equation}
and therefore
\begin{equation}
    \langle t_f\rangle_{\rm th}
        = \frac{x_0\,x_{\rm th}\bigl(x_{\rm th}^{\alpha+1}-x_0^{\alpha+1}\bigr)}
               {(\alpha+1)(\alpha+2)\,(x_{\rm th}-x_0)}.
    \label{eq:MFPT_th_explicit}
\end{equation}
The MFPT is finite for any $x_{\rm th}<\infty$ and any $\alpha\geq0$, in contrast to the free sluggish walk, for which it diverges. Writing $u=x_{\rm th}/x_0>1$, Eq.~\eqref{eq:MFPT_th_explicit} reads $\langle t_f\rangle_{\rm th}=x_0^{\alpha+2}\,u\,g(u)/[(\alpha+1)(\alpha+2)]$ with $g(u)=(u^{\alpha+1}-1)/(u-1)=(\alpha+1)\int_0^1[1+t(u-1)]^{\alpha}\,dt$. For $\alpha\ge0$ the function $g$ is positive and non-decreasing in $u$ (constant for $\alpha=0$), so $u\,g(u)$ is strictly increasing and $\langle t_f\rangle_{\rm th}$ increases monotonically with the threshold. It tends to $x_0^{\alpha+2}/(\alpha+2)$ as $x_{\rm th}\to x_0^+$, where both $\langle\tau_{\rm ex}\rangle$ and $\pi(x_0)$ vanish linearly in $x_{\rm th}-x_0$ so that their ratio remains finite, and it grows as $x_0x_{\rm th}^{\alpha+1}/[(\alpha+1)(\alpha+2)]$ for $x_{\rm th}\gg x_0$, where confinement is progressively lost. Lowering the threshold therefore always shortens the search, although the resetting frequency $1/\langle\tau_{\rm ex}\rangle$ diverges as $x_{\rm th}\to x_0^+$. The splitting probability~\eqref{eq:splitting} depends on $x_{\rm th}$ but not on $\alpha$, so the $\alpha$-dependence of the MFPT enters through $\langle\tau_{\rm ex}\rangle$. For the parameters of Fig.~\ref{fig:5} ($x_0=1$ and $x_{\rm th}\geq5$) a more sluggish walker, which moves more slowly away from the origin since $D(x)=x^{-\alpha}<1$ for $x>1$, takes longer excursions and hence a longer time to reach the target. As in the Poissonian resetting scenario, the  first-passage time density also develops an exponential tail at long times,
\begin{equation}
    F_{\rm th}(x_0,t) \sim A_{\rm th}\,e^{s_{\rm th}t}, \qquad t\to\infty,
    \label{eq:Fth_tail}
\end{equation}
consistent with the finite MFPT established above. Unlike the Poissonian pole $s_0$~\eqref{eq:pole}, the decay rate $s_{\rm th}<0$ is not obtained from a closed-form expression, but it is fixed by a transcendental equation that follows from the renewal structure. Let $\widetilde{g}_{0}(\xi,s)$ and $\widetilde{g}_{\rm th}(\xi,s)$ be the Laplace transforms of the probability densities that an excursion starting at $\xi\in(0,x_{\rm th})$ ends at the origin and at $x_{\rm th}$, respectively. Both solve $s\,\widetilde{g}=\xi^{-\alpha}\partial_\xi^2\widetilde{g}$, with $\widetilde{g}_0=1$, $\widetilde{g}_{\rm th}=0$ at $\xi=0$ and $\widetilde{g}_0=0$, $\widetilde{g}_{\rm th}=1$ at $\xi=x_{\rm th}$. Since every excursion starts at $x_0$ after a reset, the FPT density satisfies $\widetilde{F}_{\rm th}(x_0,s)=\widetilde{g}_0(x_0,s)/[1-\widetilde{g}_{\rm th}(x_0,s)]$. Solving the two linear problems with the Bessel functions used in Eq.~\eqref{eq:Q0_closed} gives
\begin{equation}
    \widetilde{F}_{\rm th}(x_0,s)
        = \frac{\Gamma(1-\mu)\,(y_0/2)^{\mu}\,
          \bigl[I_{-\mu}(y_0)I_{\mu}(y_{\rm th})-I_{\mu}(y_0)I_{-\mu}(y_{\rm th})\bigr]}
          {I_{\mu}(y_{\rm th})-(x_0/x_{\rm th})^{1/2}\,I_{\mu}(y_0)},
    \quad y_{0,{\rm th}}=\frac{2\sqrt{s}\,x_{0,{\rm th}}^{(\alpha+2)/2}}{\alpha+2},
    \label{eq:Fth_laplace}
\end{equation}
where $I_{\pm\mu}$ are modified Bessel functions of the first kind~\citep{AbramowitzStegun}. This expression satisfies $\widetilde{F}_{\rm th}\to1$ as $s\to0$, and its first-order expansion in $s$ reproduces the MFPT~\eqref{eq:MFPT_th_explicit}. The decay rate is $s_{\rm th}=-\kappa^2$, where $\kappa$ is the smallest positive root of the vanishing of the denominator in Eq.~\eqref{eq:Fth_laplace}, which for $s<0$ reads
\begin{equation}
    J_{\mu}\!\left(\frac{2\kappa\,x_{\rm th}^{(\alpha+2)/2}}{\alpha+2}\right)
        = \left(\frac{x_0}{x_{\rm th}}\right)^{1/2}
          J_{\mu}\!\left(\frac{2\kappa\,x_0^{(\alpha+2)/2}}{\alpha+2}\right),
    \label{eq:sth_eq}
\end{equation}
with $J_\mu$ the Bessel function of the first kind. For $\alpha=0$ ($\mu=1/2$) this reduces to $\sin(\kappa x_{\rm th})=\sin(\kappa x_0)$, so that $s_{\rm th}=-\pi^2/(x_{\rm th}+x_0)^2$. The amplitude is the residue of Eq.~\eqref{eq:Fth_laplace} at the pole, $A_{\rm th}=\lim_{s\to s_{\rm th}}(s-s_{\rm th})\widetilde{F}_{\rm th}(x_0,s)$, so that $s_{\rm th}$ and $A_{\rm th}$ in the dashed lines of Fig.~\ref{fig:5}, and the exact solid lines, involve no fitted parameter.

Because the splitting probability~\eqref{eq:splitting} is independent of $\alpha$, any dependence of the first-passage time density on the sluggishness parameter must generally enter through the mean excursion duration $\langle \tau_{\rm ex} \rangle$, and this is reflected in Fig.~\ref{fig:5}. At fixed $\alpha$, reducing the threshold $x_{\rm th}$ steepens the exponential decay (Fig.~\ref{fig:5}, left panel), since a walker confined to a narrower interval reaches the boundary more quickly and is therefore reset, or absorbed at the origin, on a shorter timescale. But at fixed $x_{\rm th}$, increasing $\alpha$ flattens the decay (Fig.~\ref{fig:5}, right panel), consistent with the larger mean excursion duration~\eqref{eq:tau_exit} for these parameters. The finite MFPT~\eqref{eq:MFPT_th} is therefore tuned by both the confinement geometry and the sluggishness of the dynamics.

\section{Occupation statistics}\label{sec:6}
\subsection{Distribution of occupation times}
\smallskip
Another key observable, alternative to first-passage times,  characterising the  statistical properties of a stochastic process is the {\em occupation time} on the positive half-axis, defined as the total time the walker spends at positive positions during an observation window of duration $T$~\citep{Levy_1940, Lamperti_1958,Majumdar_2005,sarmiento}
\begin{equation}
    t_+ = \int_0^T \Theta(x(s))\,ds,
    \qquad 0 \leq t_+ \leq T,
    \label{eq:occ_def}
\end{equation}
where $\Theta(\cdot)$ is the Heaviside step function. The scaled fraction $\xi = t_+/T$ takes values in $[0,1]$, and its probability density $p_+(\xi)$ encodes the tendency of the walker to remain on one side. For a simple symmetric Brownian motion, Lévy's arcsine law $p_+(\xi) = 1/(\pi\sqrt{\xi(1-\xi)})$, which is U-shaped and diverges at both endpoints, reflecting the well-known tendency of Brownian paths to spend the bulk of their time on one side of the origin~\citep{Levy_1940, Feller_1950}.

For the sluggish random walk without resetting, the successive returns to the origin define a renewal process whose inter-renewal times inherit the power-law tail $\psi(\tau) \sim \tau^{-1-\mu}$ from the first-passage result~\eqref{eq:F0_tail}, with persistence exponent $\mu = 1/(\alpha+2)$~\citep{del2025generalized}. Since positive and negative excursions are equally likely by symmetry, Lamperti's theorem~\citep{Lamperti_1958, Godreche_2001} applies directly and gives the exact scaling form
\begin{equation}
    p_+(t_+\mid T) = \frac{1}{T}\,f_+\!\left(\frac{t_+}{T}\right),
    \label{eq:occ_scaling}
\end{equation}
with the scaling function~\citep{del2025generalized}
\begin{equation}
    f_+(\xi)
        = \frac{\sin(\pi\mu)}{\pi}
          \frac{\bigl[\xi(1-\xi)\bigr]^{\mu-1}}
               {\xi^{2\mu} + 2\cos(\pi\mu)\bigl[\xi(1-\xi)\bigr]^\mu
                + (1-\xi)^{2\mu}},
    \qquad 0 \leq \xi \leq 1.
    \label{eq:f_plus}
\end{equation}
Equation~\eqref{eq:f_plus} is the Lamperti distribution with parameter $\mu$. For $\alpha = 0$, $\mu = 1/2$ and~\eqref{eq:f_plus} it reduces  to Lévy's arcsine law~\citep{Levy_1940}. For $\alpha > 0$ and $\mu < 1/2$ the distribution remains symmetric around $\xi = 1/2$, with $\langle\xi\rangle = 1/2$ for all $\alpha$, but diverges more strongly at both endpoints as $f_+(\xi) \sim \xi^{\mu-1}$ as $\xi \to 0$ and $(1-\xi)^{\mu-1}$ as $\xi \to 1$. This sharpening of the U-shape reflects the increasing tendency of the sluggish walker to remain on one side of the origin for extended periods as $\alpha$ grows~\citep{del2025generalized}.

\subsubsection{Occupation time of the sluggish random walk with Poissonian resetting}
Poissonian resetting at rate $r$ to the origin preserves a renewal structure, since the walker restarts afresh from the origin at every reset. It does, however, interrupt the excursions and cut off their power-law tail, so that the Lamperti form of the free walk is lost beyond the timescale $1/r$~\citep{denHollander_2019}. The renewal property allows the occupation time distribution to be characterised exactly at the level of its double Laplace transform. To this end, we define the generating function
\begin{equation}
    \phi(x_0, T \mid p)
        = \bigl\langle e^{-p\,t_+}\bigr\rangle_{x_0}
        = \left\langle \exp\!\left(
          -p\int_0^T \Theta\bigl(x(t)\bigr)\,dt\right)\right\rangle_{x_0},
    \label{eq:gen_func}
\end{equation}
where the average is taken over all trajectories starting at $x(0) = x_0$. In the presence of Poissonian resetting, $\phi$ satisfies the backward Feynman--Kac equation~\citep{Majumdar_2005, del2025generalized}
\begin{equation}
    \partial_T\phi
        = \frac{1}{|x_0|^\alpha}\,\partial_{x_0}^2\phi
          - p\,\Theta(x_0)\,\phi
          - r\,\phi
          + r\,\phi(0,T\mid p),
    \label{eq:FK_reset}
\end{equation}
with initial condition $\phi(x_0, 0\mid p) = 1$, which follows from the fact that $t_+ = 0$ at $T = 0$.  The term $|x_0|^{-\alpha}\partial_{x_0}^2\phi$ is the backward generator of the free sluggish process, $-r\phi$ accounts for loss of probability due to resetting, and $r\,\phi(0,T\mid p)$ reinjects it at the origin.

Taking the Laplace transform $\widetilde{\phi}(x_0,s,p) = \int_0^\infty e^{-sT}\phi(x_0,T\mid p)\,dT$ of~\eqref{eq:FK_reset} and using the initial condition together with the identity $\mathcal{L}[\partial_T\phi] = s\widetilde{\phi} - 1$ reduces the partial differential equation~\eqref{eq:FK_reset} to the ordinary differential equation
\begin{equation}
    \frac{1}{|x_0|^\alpha}
        \frac{d^2\widetilde{\phi}}{dx_0^2}
        - \bigl[p\,\Theta(x_0) + s + r\bigr]\,\widetilde{\phi}
        = -\Lambda,
    \label{eq:ODE_reset}
\end{equation}
where $\Lambda \equiv 1 + r\,\widetilde{\phi}(0,s,p)$ collects the constant terms on the right-hand side.

The effective decay rate in~\eqref{eq:ODE_reset} differs between the two half-lines. On the positive side $x_0 > 0$ the indicator $\Theta(x_0)=1$ gives a total rate $s+r+p$, while on the negative side $\Theta(x_0) = 0$ gives simply $s+r$.  We solve~\eqref{eq:ODE_reset} on each half-line separately and match at the origin.  On the positive side, the particular solution is $\Lambda/(s+r+p)$.  Setting
\begin{equation}
    z_+ = \frac{2\sqrt{s+r+p}}{\alpha+2}\,x_0^{(\alpha+2)/2},
    \qquad
    \widetilde{\phi}(x_0) = x_0^{1/2} f(z_+) + \frac{\Lambda}{s+r+p},
    \label{eq:cov_pos}
\end{equation}
transforms the homogeneous part of~\eqref{eq:ODE_reset} into the modified Bessel equation of order $\mu = 1/(\alpha+2)$
\begin{equation}
    \frac{d^2f}{dz_+^2}
        + \frac{1}{z_+}\frac{df}{dz_+}
        - \left(1 + \frac{\mu^2}{z_+^2}\right)f = 0.
    \label{eq:bessel_pos}
\end{equation}
Since $I_\mu(z_+)$ grows exponentially as $x_0 \to \infty$, normalisability requires discarding it, leaving $f(z_+) = A_+ K_\mu(z_+)$.  The full solution on the positive side is therefore
\begin{equation}
    \widetilde{\phi}(x_0, s, p)
        = A_+\,x_0^{1/2}\,K_\mu\!\left(
          \frac{2\sqrt{s+r+p}}{\alpha+2}\,x_0^{(\alpha+2)/2}\right)
          + \frac{\Lambda}{s+r+p},
    \qquad x_0 > 0.
    \label{eq:sol_pos}
\end{equation}
An identical argument on the negative side, with $z_- = \frac{2\sqrt{s+r}}{\alpha+2}|x_0|^{(\alpha+2)/2}$, gives
\begin{equation}
    \widetilde{\phi}(x_0, s, p)
        = A_-\,|x_0|^{1/2}\,K_\mu\!\left(
          \frac{2\sqrt{s+r}}{\alpha+2}\,|x_0|^{(\alpha+2)/2}\right)
          + \frac{\Lambda}{s+r},
    \qquad x_0 < 0.
    \label{eq:sol_neg}
\end{equation}
The constants $A_+$ and $A_-$ are fixed by two conditions at $x_0 = 0$. The first is continuity of $\widetilde{\phi}$.  Using the small-argument behaviour $K_\mu(z) \sim \frac{\Gamma(\mu)}{2}(z/2)^{-\mu}$ as $z \to 0$, both solutions approach finite limits as $x_0 \to 0$, and equating them gives
\begin{equation}
    A_+\,\frac{\Gamma(\mu)}{2}
        \left(\frac{\sqrt{s+r+p}}{\alpha+2}\right)^{-\mu}
        + \frac{\Lambda}{s+r+p}
        = A_-\,\frac{\Gamma(\mu)}{2}
        \left(\frac{\sqrt{s+r}}{\alpha+2}\right)^{-\mu}
        + \frac{\Lambda}{s+r}.
    \label{eq:continuity}
\end{equation}
The second condition follows from integrating Eq.~\eqref{eq:ODE_reset}, written as $\widetilde{\phi}''=|x_0|^{\alpha}\{[p\,\Theta(x_0)+s+r]\widetilde{\phi}-\Lambda\}$, across an infinitesimal neighbourhood of the origin. The right-hand side is bounded there and is multiplied by $|x_0|^{\alpha}$, so its integral vanishes with the width of the neighbourhood, and $d\widetilde{\phi}/dx_0$ is continuous at $x_0=0$. Writing $K_\mu=\frac{\pi}{2\sin(\pi\mu)}\,[I_{-\mu}-I_{\mu}]$, the term linear in $x_0$ arises only from $I_\mu$, which gives $d\widetilde{\phi}/dx_0|_{0^+}=A_+\frac{\Gamma(-\mu)}{2}\bigl(\frac{\sqrt{s+r+p}}{\alpha+2}\bigr)^{\mu}$ and $d\widetilde{\phi}/dx_0|_{0^-}=-A_-\frac{\Gamma(-\mu)}{2}\bigl(\frac{\sqrt{s+r}}{\alpha+2}\bigr)^{\mu}$. Continuity of the slope therefore reads
\begin{equation}
    A_+\,(s+r+p)^{\mu/2} = -A_-\,(s+r)^{\mu/2}.
    \label{eq:slope_phi}
\end{equation}
Solving the linear system formed by~\eqref{eq:continuity} and~\eqref{eq:slope_phi} for $A_+$ and $A_-$ and substituting back into~\eqref{eq:sol_pos} evaluated as $x_0 \to 0^+$ gives
\begin{equation}
    \widetilde{\phi}(0,s,p)
        = \Lambda\,\frac{a^{\mu-1}+c^{\mu-1}}{a^{\mu}+c^{\mu}},
    \qquad a=s+r+p,\quad c=s+r.
    \label{eq:phi0_Lambda}
\end{equation}
Invoking the definition $\Lambda = 1 + r\,\widetilde{\phi}(0,s,p)$ gives a self-consistency equation that determines $\Lambda$ uniquely and yields the exact double Laplace transform
\begin{equation}
    \widetilde{\phi}(0,\,s,\,p)
        = \frac{a^{\mu-1}+c^{\mu-1}}
               {(s+p)\,a^{\mu-1}+s\,c^{\mu-1}},
    \label{eq:phi_result}
\end{equation}
where $\mu = 1/(\alpha+2)$.  Setting $p=0$ gives $\widetilde{\phi}(0,s,0)=1/s$, as required by normalisation. For $r=0$, Eq.~\eqref{eq:phi_result} reduces to $[(s+p)^{\mu-1}+s^{\mu-1}]/[(s+p)^{\mu}+s^{\mu}]$, which is the double Laplace transform of the Lamperti distribution~\eqref{eq:f_plus}~\citep{Godreche_2001}. For $p\to\infty$ it vanishes, consistent with $t_+>0$ with probability one for a walker that starts at the origin.

\begin{figure}[h]
    \centering
    \includegraphics[width=0.875\textwidth, trim=0cm 0cm 0cm 0cm, clip]{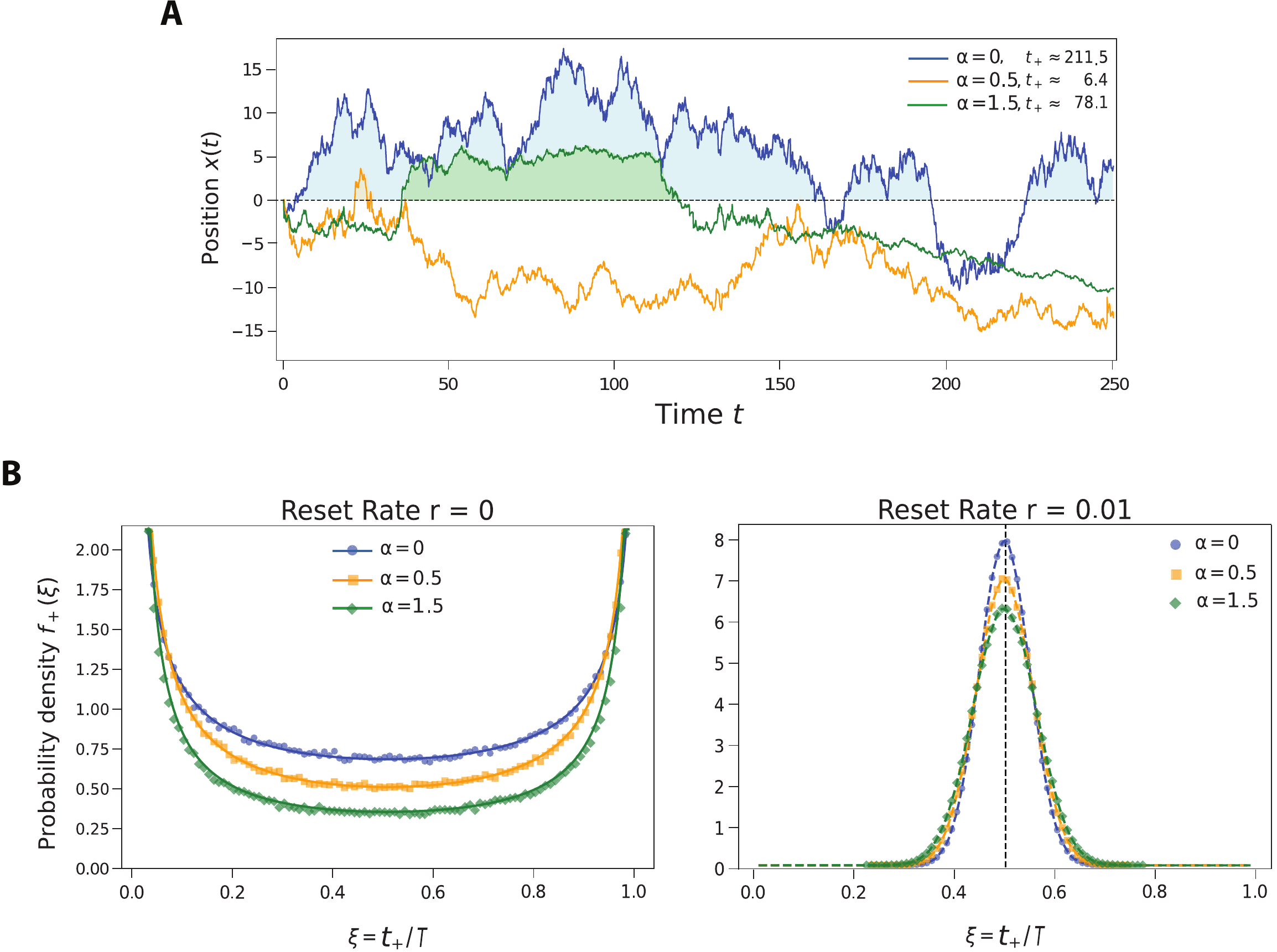}
    \caption{Occupation-time statistics of the sluggish random walk under Poissonian resetting. \textbf{(A)} Example trajectories $x(t)$ at a fixed resetting rate $r = 0.01$ for $\alpha \in \{0,\, 0.5,\, 1.5\}$ (see legends), illustrating the time $t_+$ spent on the positive half-axis (shaded), which defines the occupation fraction $\xi = t_+/T$. \  \textbf{(B)} Probability density of the occupation fraction $\xi = t_+/T$ for sluggishness parameters $\alpha \in \{0,\,0.5,\,1.5\}$ (see legend) and reset rates $r = 0$ (no resetting, left panel) and $r = 0.01$ (Poissonian resetting, right panel). Symbols are obtained from $N = 10^6$ independent simulations with observation times $T = 10^4$, and lines are given by the exact Lamperti distribution $f_+(\xi)$~\eqref{eq:f_plus} for $r=0$, and by kernel density estimates for $r=0.01$.}
    \label{fig:6}
\end{figure}

The dimensionless product $rT$ controls the crossover between two limiting occupation-time regimes. When $rT \ll 1$, resets are rare over the observation window, and the distribution of $\xi=t_+/T$ approaches the free-sluggish Lamperti form~\eqref{eq:f_plus}, retaining its pronounced weight near $\xi=0$ and $\xi=1$. In contrast, when $rT \gg 1$, the walker undergoes many resets to the origin before the observation time $T$ is reached. Since each reset restarts the walker afresh at the origin, the observation window is made of about $rT$ independent segments of exponentially distributed durations $t_i$ with mean $1/r$. Within each segment the occupation fraction $\xi_i$ follows the free Lamperti law~\eqref{eq:f_plus}, with mean $1/2$ and variance $(1-\mu)/4$, so that $\xi-1/2=\sum_i t_i(\xi_i-1/2)/\sum_i t_i$. Frequent resetting thus prevents trajectories from spending almost the entire observation interval on only one side of the origin. The central limit theorem gives a Gaussian distribution centred at $1/2$,
\begin{equation}
    p_+(\xi)\simeq\sqrt{\frac{rT}{\pi(1-\mu)}}\,
    \exp\!\left[-\frac{rT\,(\xi-1/2)^2}{1-\mu}\right],
    \qquad
    \mathrm{Var}(\xi)=\frac{1-\mu}{2rT}=\frac{\alpha+1}{2(\alpha+2)\,rT}.
    \label{eq:gaussian_occ}
\end{equation}
The same variance follows from expanding Eq.~\eqref{eq:phi_result} to second order in $p$, which gives $\langle t_+\rangle=T/2$ and $\mathrm{Var}(t_+)\simeq(1-\mu)T/(2r)$ at large $T$. The U-shaped divergences near $\xi=0$ and $\xi=1$ of the free Lamperti law are replaced by a Gaussian peak whose width decreases as $(rT)^{-1/2}$ and, at fixed $rT$, grows with $\alpha$ through the factor $1-\mu$. This tendency is already evident in Fig.~\ref{fig:6}(B), where the distributions at $r=0.01$ are noticeably narrower than their $r=0$ counterparts and exhibit reduced accumulation near the occupation-time boundaries.

\subsubsection{Occupation time of the sluggish random walk with threshold resetting}
Threshold resetting gives the occupation time a renewal structure. The walker starts at and is reset to $x_0>0$, with thresholds at $\pm x_{\rm th}$. Each excursion from $x_0$ ends when the walker first reaches $\pm x_{\rm th}$, at which point it resets to $x_0$ and a new statistically independent excursion begins.  The successive departures from $x_0$ are therefore the renewal points of the process, and the inter-renewal time $\tau$ is the first-exit time from the interval $(-x_{\rm th},x_{\rm th})$ starting at $x_0$, irrespective of the threshold at which it ends.

Unlike the free sluggish walk, the inter-renewal times here have a finite mean, since the walker is confined to the bounded interval $(-x_{\rm th}, x_{\rm th})$ and all moments of the FPT from $x_0$ to $\pm x_{\rm th}$ are finite; the power-law tail $\psi(\tau)\sim\tau^{-1-\mu}$ of the free process is destroyed by the confinement. The condition for Lamperti's theorem, infinite mean inter-renewal time, is therefore not satisfied, and the long-time occupation statistics are governed instead by the renewal theorem~\citep{Feller_1950}.  For large $T$, the law of large numbers applied to the renewal sequence gives a deterministic limit.
\begin{equation}
    \xi
        = \frac{t_+}{T}
        \;\xrightarrow{T\to\infty}\;
        \xi^*
        \equiv
        \frac{\langle\eta\,\tau\rangle}
             {\langle \tau\rangle},
    \label{eq:xi_star}
\end{equation}
where $\tau$ is the duration of a single excursion and $0\leq \eta\leq 1$ is the fraction of time that excursion spent on the positive half-axis. Both expectations are finite, so $\xi^*$ is well defined. They follow from the mean time spent by an excursion at position $y$, which is $|y|^{\alpha}g(y,x_0)$, where $g$ is the Green function of $-d^2/dy^2$ on $(-x_{\rm th},x_{\rm th})$ with Dirichlet conditions at the thresholds. Integrating over the full interval gives the mean excursion duration
\begin{equation}
    \langle\tau\rangle = \frac{x_{\rm th}^{\alpha+2}-x_0^{\alpha+2}}{(\alpha+1)(\alpha+2)},
\end{equation}
and integrating over the negative half-axis gives 
\begin{equation}
    \langle(1-\eta)\tau\rangle = \frac{(x_{\rm th}-x_0)\,x_{\rm th}^{\alpha+1}}{2(\alpha+1)(\alpha+2)}.
\end{equation} 
Therefore
\begin{equation}
    \xi^*
        = 1-\frac{1-\rho}{2\left(1-\rho^{\alpha+2}\right)},
    \qquad \rho=\frac{x_0}{x_{\rm th}}.
    \label{eq:xi_star_closed}
\end{equation}
Hence $\xi^*$ depends only on the confinement ratio $\rho$ and on $\alpha$, and not on $T$.

The value of $\xi^*$ reflects a competition between the initial bias towards the positive side, since the walker always restarts from $x_0 > 0$, and the symmetric diffusion during each excursion. Equation~\eqref{eq:xi_star_closed} decreases monotonically with $x_{\rm th}$, from $\xi^*\to1-\mu/2=(2\alpha+3)/(2\alpha+4)$ for $x_{\rm th}\to x_0$ to $\xi^*\to1/2$ for $x_{\rm th}\to\infty$, so that $1/2<\xi^*<1-\mu/2$. The upper bound equals $3/4$, $4/5$ and $6/7$ for $\alpha=0$, $0.5$ and $1.5$, and it lies strictly below one because, even when $x_{\rm th}$ is close to $x_0$ and most excursions end quickly at the upper threshold, the rare excursions that reach the negative half-axis last much longer. At fixed $\rho$, $\xi^*$ increases with $\alpha$, since $\rho^{\alpha+2}$ in Eq.~\eqref{eq:xi_star_closed} decreases with $\alpha$.

\begin{figure}
    \centering
    \includegraphics[width=1\textwidth, trim=0cm 0cm 0cm 0cm, clip]{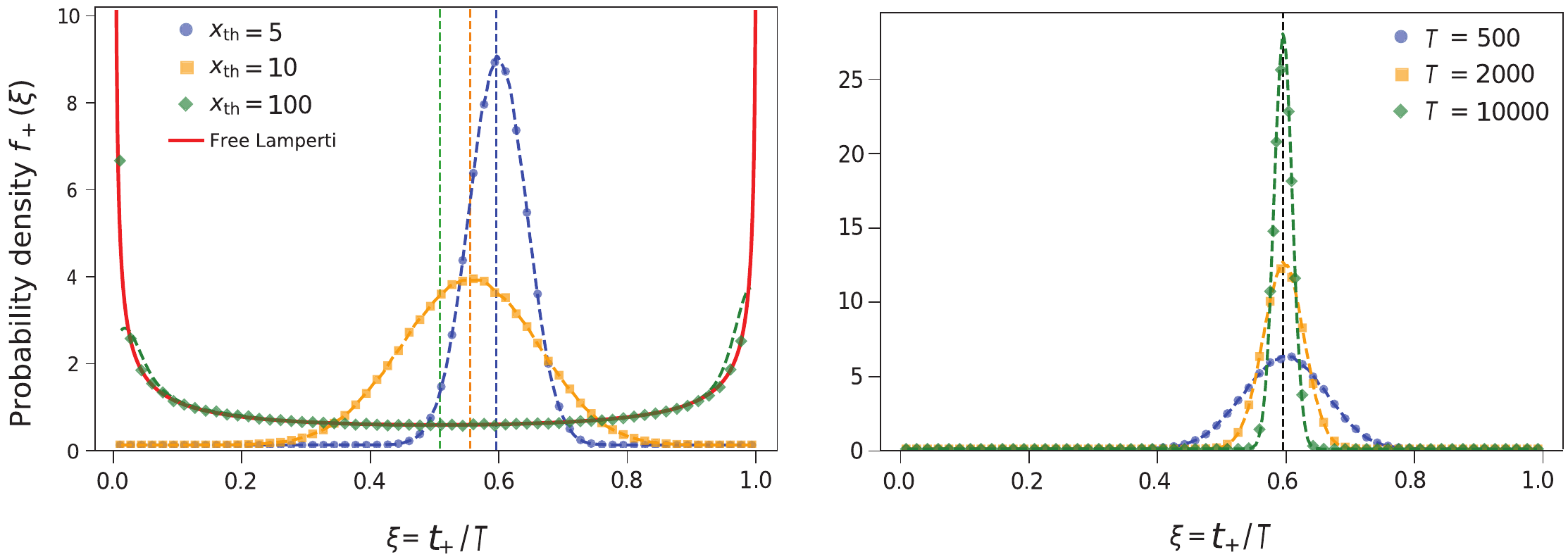}
    \caption{Distribution of the occupation fraction $\xi = t_+/T$ for the sluggish walk with threshold resetting at $\alpha = 0.5$, $x_0 = 1$, obtained from $N = 10^6$ independent simulations. Left panel, fixed observation time $T = 10^4$ with thresholds $x_{\rm th} \in \{5,\,10,\,100\}$ (see legend), showing that $\xi^*$ shifts systematically towards $1/2$ as $x_{\rm th}$ increases. Red curve shows the free Lamperti distribution $f_+(\xi)$~\eqref{eq:f_plus}, recovered in the limit $x_{\rm th} \to \infty$ at fixed $T$. Right panel, fixed threshold $x_{\rm th} = 5$ with observation times $T \in \{500,\,2000,\,10^4\}$ (see legend), showing the progressive concentration of $\xi$ around the deterministic limit $\xi^*$~\eqref{eq:xi_star}. Symbols are histogram estimates of the simulated distribution, and the coloured dashed curves are Gaussian kernel density estimates of the same data. The analytical results are the Lamperti law (red curve) and $\xi^*$~\eqref{eq:xi_star} (vertical lines).}
    \label{fig:7}
\end{figure}

The limits $T\to\infty$ and $x_{\rm th}\to\infty$ do not commute. Taking $x_{\rm th}\to\infty$ after $T\to\infty$ gives $\xi\to\xi^*\to1/2$. The free Lamperti law~\eqref{eq:f_plus} is instead recovered when $x_{\rm th}\to\infty$ at fixed $T$, because the walker, which travels a distance of order $T^{\mu}$ in time $T$, then never reaches the thresholds. The crossover is controlled by $T/\langle\tau\rangle$, the typical number of cycles completed during the observation window, with $\langle\tau\rangle\sim x_{\rm th}^{\alpha+2}$. For $T\ll\langle\tau\rangle$ the occupation fraction follows the free Lamperti law, while for $T\gg\langle\tau\rangle$ it concentrates around $\xi^*$ with fluctuations of order $(T/\langle\tau\rangle)^{-1/2}$. As shown in Fig.~\ref{fig:7} (left panel), increasing $x_{\rm th}$ at fixed $T$ shifts the histogram of $\xi$ towards $1/2$ and towards the free Lamperti curve.

This behaviour stands in sharp contrast to both the free sluggish walk, where $\xi$ follows the U-shaped Lamperti distribution~\eqref{eq:f_plus}, and to Poissonian resetting to the origin, where the distribution is a Gaussian centred at $1/2$ with a width that decreases as $(rT)^{-1/2}$, see Eq.~\eqref{eq:gaussian_occ}. Threshold resetting to $x_0>0$ produces a third distinct regime. At long times ($T\gg\langle\tau\rangle$), the distribution of $\xi$ collapses onto the deterministic value $\xi^*$ defined in~\eqref{eq:xi_star}, with $1/2 < \xi^* < 1-\mu/2$ determined by $x_0/x_{\rm th}$ and $\alpha$ through Eq.~\eqref{eq:xi_star_closed}. This is confirmed in Fig.~\ref{fig:7} (right panel), where the histogram of $\xi$ visibly sharpens around $\xi^*$ as the observation time $T$ grows at fixed $x_{\rm th}=5$.

\section{Cost of resetting sluggish random walks}\label{sec:7}
\smallskip
\paragraph{} 
Resetting drives the sluggish random walk into a nonequilibrium stationary state by breaking detailed balance. Thus the rate of entropy production associated with sluggish random walks with non-zero resetting rates  is strictly positive, as dictated by the second law of stochastic thermodynamics~\citep{Seifert_2012}. The contribution of resetting to the total entropy production  $\dot S_{\rm reset}$, need not itself be positive. It is negative for ordinary diffusion, reflecting a net erasure of positional information each time the particle is returned to $x_0$~\citep{Fuchs_2016, Gupta_2022}.

An alternative perspective on the cost of maintaining the nonequilibrium stationary state induced by resetting assigns a physical, phenomenological cost to each reset, such as the time and/or energy required to return the particle to the origin~\citep{Sunil_2023, Pal_2020}. We refer to this quantity, a function $c(|x|)$ of the position from which the reset occurs, as the {\em cost per reset}, and to its average per unit time in the stationary state, $\langle\dot{\mathcal{C}}\rangle_{\rm st}=r\langle c\rangle_{\rm st}$, as the {\em cost rate}. For ordinary diffusion with Poissonian resetting, a cost per reset that scales with the displacement as $c \sim |x-x_0|^\beta$ gives a cost rate $\langle\dot{\mathcal{C}}\rangle_{\rm st}\sim r^{1-\beta/2}$, since the stationary distribution has width $r^{-1/2}$. The cost rate therefore vanishes as $r \to 0$ for $\beta < 2$, remains finite for $\beta = 2$, and diverges for $\beta > 2$. This differs from the transition of the mean total cost accumulated until first passage to a target, which occurs at $\beta=1$~\citep{Sunil_2023}.

The optimal resetting rate minimising the mean completion time, defined in Ref.~\citep{Sunil_2023} as the sum of the first-passage time $t_f$ and a total cost $t_\mathcal{C}$ normalised to the dimension of time, depends on the form of the cost function and in general differs from the rate that minimises the first-passage time alone. Of the cost functions considered below, only the linear cost $c(|x|)=|x|$, proportional to the time needed to return at constant speed, carries the dimension of time directly. The power-law and exponential costs are better interpreted as more general resource penalties rather than literal contributions to a completion time. Here we study the stationary cost rate, for which an optimal resetting rate also emerges for the exponential cost (Sec.~\ref{sec:poisson_avg_cost}).

In what follows, we study the resetting costs using the same three families of cost function, linear, power-law, and exponential in the displacement, for both Poissonian resetting (Sec.~\ref{sec:poisson_cost}) and threshold resetting (Sec.~\ref{sec:threshold_cost}), so that the two resetting mechanisms can be compared directly under identical cost functions.

\subsection{Poissonian Resetting case}\label{sec:poisson_cost}
\smallskip

\subsubsection{Entropy production in the nonequilibrium stationary state}
\label{sec:poisson_cost_entropy}
A natural measure of the thermodynamic cost of the resetting process is the entropy production rate, which vanishes in equilibrium and is strictly positive in any nonequilibrium stationary state~\citep{Seifert_2012}. For a system driven into a continuous-time nonequilibrium stationary state by stochastic resetting, the rate of entropy production associated with the resetting events is given by Eq.~(12) of~\citep{Fuchs_2016} (see also~\citep{Busiello_2020}),
\begin{equation}
    \dot{S}_{\mathrm{reset}}
        = r \int_{-\infty}^{\infty} P_{\mathrm{st}}(x)
          \ln\!\left( \frac{P_{\mathrm{st}}(x)}{P_{\mathrm{st}}(x_0)} \right) dx,
    \label{eq:entropy}
\end{equation}
where $x_0 = 0$ is the resetting location. This expression involves only the stationary distribution and the resetting location. Each reset event removes probability from position $x$ and teleports it at the origin, and the logarithmic ratio measures the (Shannon) informational cost of this transition~\citep{Fuchs_2016}.
For the diffusive case $\alpha = 0$, the stationary distribution reduces to the bilateral exponential $P_{\mathrm{st}}(x) = (\sqrt{r}/2) e^{-\sqrt{r}|x|}$ [see Eq.~\eqref{eqn:svapdr}]. Since $P_{\mathrm{st}}(x)/P_{\mathrm{st}}(0) = e^{-\sqrt{r}|x|}$, Eq.~\eqref{eq:entropy} reduces to

\begin{equation}
    \dot{S}_{\mathrm{reset}}\big|_{\alpha=0}
        = -r^{3/2}\int_{-\infty}^{\infty} |x|\,P_{\mathrm{st}}(x)\,dx
        = -r^{3/2}\,\langle |x| \rangle
        = -r,
    \label{eq:entropy_alpha0}
\end{equation}
where we have used $\langle |x| \rangle = 1/\sqrt{r}$ for the bilateral exponential distribution. This result is negative, consistent with the fact that resetting concentrates probability near the origin, and its magnitude grows linearly with~$r$. For $\alpha>0$, the stationary distribution~\eqref{eqn:bessel} vanishes at the origin, $P_{\mathrm{st}}(0)=0$, since the time spent around $x$ is proportional to $1/D(x)=|x|^{\alpha}$. The ratio $P_{\mathrm{st}}(x)/P_{\mathrm{st}}(x_0)$ then diverges for every $x$ with $P_{\mathrm{st}}(x)>0$, and Eq.~\eqref{eq:entropy} gives $\dot S_{\rm reset}=+\infty$ for all $\alpha>0$ and $r>0$. This divergence follows from the singular diffusivity at the origin in the continuum limit. On the lattice, where the hopping probabilities~\eqref{eqn:hop_prob} are bounded, $P_{\rm st}(x_0)$ is finite and so is $\dot S_{\rm reset}$. Close to the reset site the free chain obeys detailed balance, so that $P_{\rm st}(n)\propto|n|^{\alpha}+2$ and $\ln[P_{\rm st}(n)/P_{\rm st}(0)]\simeq\alpha\ln|n|+O(1)$ for $1\ll|n|\ll\ell$, with $\ell=r^{-1/(\alpha+2)}$ in units of $b$. Since the stationary distribution is concentrated at $|n|\sim\ell$, Eq.~\eqref{eq:entropy} gives $\dot S_{\rm reset}\simeq r\,[\alpha\ln\ell+O(1)]=r\,[\alpha\mu\ln(1/r)+O(1)]$, which grows only logarithmically with the ratio $\ell/b$.

\subsubsection{Resetting cost}

The entropy production rate~\eqref{eq:entropy} is the cost rate~\eqref{eq:avgcost} below for the cost $c_{\rm info}(x)\equiv\ln[P_{\rm st}(x)/P_{\rm st}(x_0)]$, that is, $\dot S_{\rm reset}=\langle \dot{\mathcal{C}}\rangle_{\rm st}\big|_{c=c_{\rm info}}$. Unlike the linear, power-law and exponential costs considered next, $c_{\rm info}$ is not a positive increasing function of the reset distance: for $\alpha=0$ it equals $-\sqrt{r}\,|x|$, which is negative and decreasing, and for $\alpha>0$ it diverges in the continuum since $P_{\rm st}(x_0)=0$. On the lattice, where $P_{\rm st}(0)$ is finite, $c_{\rm info}(n)$ grows as $\alpha\ln|n|$ for $1\ll|n|\ll\ell$ and becomes negative far in the tail, where $P_{\rm st}(n)<P_{\rm st}(0)$. The phenomenological cost per reset $c(|x|)$ is instead a resource penalty, such as the time or energy needed to return the walker, and its mean remains finite for $\alpha>0$ owing to the compressed-exponential tail of $P_{\rm st}(x)$ (Sec.~\ref{sec:poisson_cost_dist}).

\subsubsection{Distribution of the cost per reset}\label{sec:poisson_cost_dist}

While the mean cost, derived below as a corollary, characterises typical behaviour, the full distribution of the cost incurred in a single reset event is key to understand nonequilibrium fluctuations and the possibility of rare but catastrophic costs~\citep{Sunil_2023, Olsen_2024}. In the steady state, the particle's position just before a reset is distributed according to $P_{\mathrm{st}}(x)$. If the cost depends only on the distance to the origin at the moment of resetting, $w = c(|x|)$, then the probability density of $w$ is obtained by a standard change of variables~\citep{Sunil_2023}
\begin{equation}
    \varrho(w) = 2\,P_{\mathrm{st}}\!\bigl(x(w)\bigr)
           \left|\frac{dx}{dw}\right|,
    \qquad w \geq w_{\min},
    \label{eq:costdist}
\end{equation}
where $x(w)$ is the inverse of $c(|x|)$ for $x > 0$ and the factor of two accounts for both signs of~$x$. The lower limit $w_{\min}\equiv c(0^+)$ is the cost of a reset from a position infinitesimally close to the origin: $w_{\min}=0$ for the linear and power-law costs below, since $c(|x|)\to 0$ as $x\to 0$, and $w_{\min}=1$ for the exponential cost $c(|x|)=e^{\lambda|x|}$, since $c(0)=1$.

Because the tail of $\varrho(w)$ for a growing cost function is controlled entirely by the large-$|x|$ behaviour of $P_{\mathrm{st}}(x)$, its behaviour near the origin plays no role here; for $\alpha>0$ this is in any case a vanishing cusp, $P_{\mathrm{st}}(x)\sim x^\alpha$ as $x\to0$, which follows from the small-argument form of $K_\nu$ in Eq.~\eqref{eqn:bessel}. Using the asymptotic form of the modified Bessel function $K_\nu(z) \sim \sqrt{\pi/(2z)}\,e^{-z}$ for large argument, the stationary distribution~\eqref{eqn:bessel} decays as
\begin{equation}
    P_{\mathrm{st}}(x)
        \sim C(\alpha, r)\,|x|^{3\alpha/4}
             \exp\!\left(-\frac{2\sqrt{r}}{\alpha+2}\,|x|^{(\alpha+2)/2}\right),
    \qquad |x| \to \infty,
    \label{eq:asympt}
\end{equation}
with $C(\alpha,r)$ a positive constant. The decay~\eqref{eq:asympt} controls both the shape of $\varrho(w)$ and the finiteness of its moments, as we now show for three families of cost functions.

\begin{itemize}
        \item {\bf Linear cost.}  For $w = |x|$ one has $x = w$ and $dx/dw = 1$, so $\varrho(w) = 2P_{\mathrm{st}}(w)$. The distribution $\varrho(w)$ therefore inherits the compressed-exponential tail of $P_{\mathrm{st}}$
\begin{equation}
    \varrho(w) \sim C(\alpha,r)\,w^{3\alpha/4}\,
              \exp\!\left(-\frac{2\sqrt{r}}{\alpha+2}\,w^{(\alpha+2)/2}\right).
\end{equation}

\item {\bf Power-law cost.}
For $w = |x|^\beta$, one has $x = w^{1/\beta}$ and $dx/dw = \beta^{-1}w^{1/\beta - 1}$, giving
\begin{equation}
    \varrho(w) \sim \frac{2}{\beta}\,w^{\frac{1}{\beta}-1}\,C(\alpha,r)\,
              w^{\frac{1}{\beta}\left(\frac{3\alpha}{4}\right)}
              \exp\!\left(-\frac{2\sqrt{r}}{\alpha+2}\,
              w^{(\alpha+2)/(2\beta)}\right),
\end{equation}
which is again of the form $\exp(-c\,w^{\zeta})$ with $\zeta=(\alpha+2)/(2\beta)$, a compressed exponential for $\beta<(\alpha+2)/2$ and a stretched exponential otherwise, so that $\varrho(w)$ has no heavy tails and all of its moments are finite regardless of $\beta$.

\item {\bf Exponential cost.} For $w = e^{\lambda|x|}$, one has $|x| = \lambda^{-1}\ln w$ and $dx/dw = 1/(\lambda w)$. Substituting into~\eqref{eq:costdist} and using~\eqref{eq:asympt},
\begin{equation}
    \varrho(w) \sim \frac{2C(\alpha,r)}{\lambda w}
              \left(\frac{\ln w}{\lambda}\right)^{\!3\alpha/4}
              \exp\!\left(-\frac{2\sqrt{r}}{\alpha+2}
              \left(\frac{\ln w}{\lambda}\right)^{\!(\alpha+2)/2}\right).
\end{equation}
For $\alpha = 0$, the exponential stationary distribution gives $\varrho(w) \sim (\sqrt{r}/\lambda)\,w^{-1-\sqrt{r}/\lambda}$, a pure power law first derived in~\citep{Sunil_2023}, whose mean diverges whenever $\lambda \geq \sqrt{r}$. For $\alpha > 0$, however, the decay is faster than any power law because the exponent involves $(\ln w)^{(\alpha+2)/2}$ with $(\alpha+2)/2 > 1$. Sluggish diffusion therefore eliminates the heavy tail that would otherwise emerge from an exponential cost function, replacing it with a distribution whose moments are all finite (see Sec.~\ref{sec:poisson_avg_cost}).
\end{itemize}

\begin{figure}
    \centering
    \includegraphics[width=1\textwidth, trim=0cm 0cm 0cm 0cm, clip]{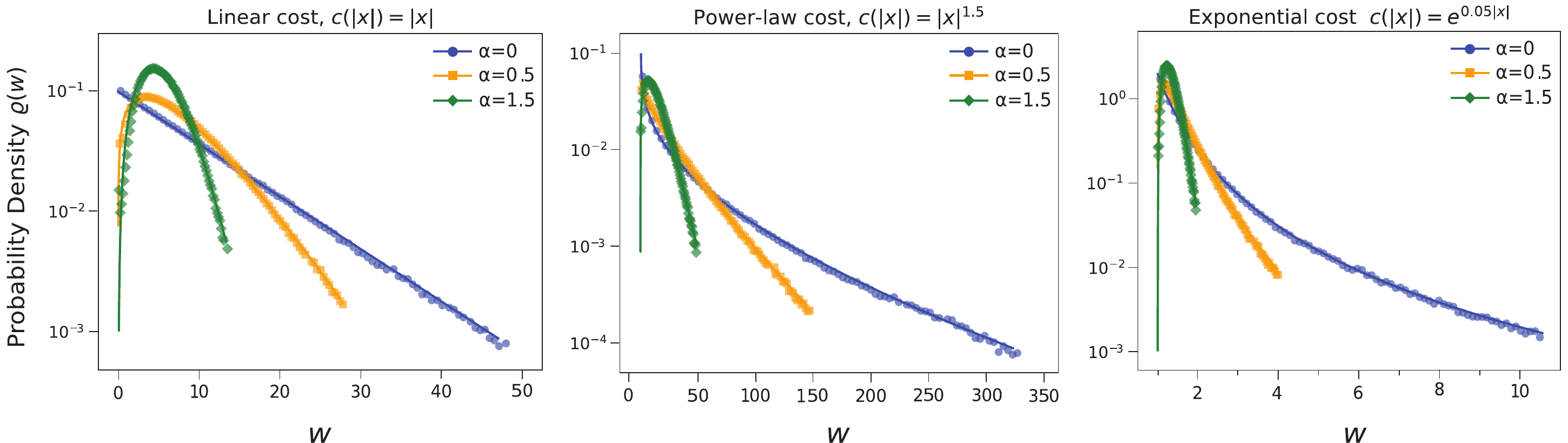}
    \caption{Probability density $\varrho(w)$ of the cost incurred in a single reset event under Poissonian resetting, Eq.~\eqref{eq:costdist}, for linear (left panel), power-law with $\beta=1.5$ (middle panel), and exponential with $\lambda=0.05$ (right panel) cost functions, and sluggishness parameters $\alpha \in \{0,\, 0.5,\, 1.5\}$ (see legend), at fixed $r=0.01$. Symbols are obtained from $N = 10^6$ independent simulations, each evolved until first reset, with the cost evaluated at the position just before the reset. Lines are given by the exact analytical result obtained from~\eqref{eq:costdist} using the stationary distribution~\eqref{eqn:bessel}.}
\label{fig:8}
\end{figure}

The suppression of large fluctuations when increasing the sluggishness parameter $\alpha$ for all the cost functions considered here is a direct consequence of the position-dependent diffusion coefficient as illustrated in Fig.~\ref{fig:8}, where the simulated histograms for all three cost functions and all values of $\alpha$ agree with the exact analytical densities~\eqref{eq:costdist}. The three cost distributions are all narrowed and shift away from the origin as $\alpha$ increases.

\subsubsection{Average cost to maintain the nonequilibrium stationary state}\label{sec:poisson_avg_cost}
Having characterised the distribution of the cost incurred in a single reset event above, we now turn to its mean. For a Poissonian resetting process with rate $r$, where each reset from position $x$ incurs a cost $c(|x|)$, the cost rate, namely the average cost per unit time needed to sustain the nonequilibrium stationary state, is~\citep{Olsen_2024, Mori_2023}
\begin{equation}
    \langle \dot{\mathcal{C}} \rangle_{\mathrm{st}}
        = r \int_{-\infty}^{\infty} c(|x|)\, P_{\mathrm{st}}(x)\, dx
        = r \int_{w_{\min}}^{\infty} w\, \varrho(w)\, dw,
    \label{eq:avgcost}
\end{equation}
where $\mathcal{C}$ denotes the  stochastic cost accumulated over the process. The second equality follows directly from the change of variables $w = c(|x|)$ used to obtain $\varrho(w)$ in~\eqref{eq:costdist}, so that $\langle \dot{\mathcal{C}} \rangle_{\mathrm{st}}$ is simply $r$ times the first moment of the cost distribution derived above. This is a direct generalisation, to sluggish dynamics, of the average cost per unit time that Sunil~\textit{et al}~\citep{Sunil_2023} propose, for normal diffusion, as a natural extension of their own results. The form of $c(\cdot)$ encodes the physical mechanism of the reset. It may represent a time penalty proportional to the return distance, an energetic expenditure, or a more general resource consumption~\citep{Sunil_2023, Pal_2020, Bodrova_2020}. Below we evaluate $\langle\dot{\mathcal{C}}\rangle_{\mathrm{st}}$ for the same three families of cost function considered above.

\paragraph{Linear cost.}
Setting $c(|x|) = |x|$ models a cost proportional to the time required to return to the origin at constant speed~\citep{Pal_2020, Sunil_2023}. Substituting into~\eqref{eq:avgcost} and applying the change of variables $z = \frac{2\sqrt{r}}{\alpha+2}\,x^{(\alpha+2)/2}$ reduces the integral to a standard Mellin transform of the modified Bessel function~\citep{Erdelyi_1954, Gradshteyn_2007}:
\begin{equation}
    \langle \dot{\mathcal{C}} \rangle_{\mathrm{st}}^{(\mathrm{lin})}
        = (\alpha+2)^{2/(\alpha+2)}\,
          \frac{\Gamma\!\left(\frac{\alpha+3}{\alpha+2}\right)}
               {\Gamma\!\left(\frac{\alpha+1}{\alpha+2}\right)}\,
          r^{(\alpha+1)/(\alpha+2)}.
    \label{eq:avgcost_lin}
\end{equation}
For $\alpha = 0$,~\eqref{eq:avgcost_lin} reduces exactly to $\langle\dot{\mathcal{C}}\rangle_{\mathrm{st}}^{(\mathrm{lin})} = \sqrt{r}$, recovering the known square-root dependence on $r$ for normal diffusion~\citep{Sunil_2023, Olsen_2024}. More generally, $\langle\dot{\mathcal{C}}\rangle_{\mathrm{st}}^{(\mathrm{lin})} \propto r^{(\alpha+1)/(\alpha+2)}$, so the sluggishness parameter also modifies the rate at which the mean cost grows with $r$, reducing to the familiar square-root law only at $\alpha=0$.

\paragraph{Power-law cost.}
For $c(|x|) = |x|^\beta$ with $\beta > 0$, the same Mellin transform as for the linear cost gives
\begin{equation}
    \langle \dot{\mathcal{C}} \rangle_{\mathrm{st}}^{(\mathrm{pow})}
        = (\alpha+2)^{2\beta/(\alpha+2)}\,
          \frac{\Gamma\!\left(\frac{\alpha+\beta+1}{\alpha+2}\right)
                \Gamma\!\left(\frac{\alpha+\beta+2}{\alpha+2}\right)}
               {\Gamma\!\left(\frac{\alpha+1}{\alpha+2}\right)}\,
          r^{1-\beta/(\alpha+2)},
    \label{eq:avgcost_pow}
\end{equation}
which reduces to Eq.~\eqref{eq:avgcost_lin} for $\beta=1$ and to $\Gamma(\beta+1)\,r^{1-\beta/2}$ for $\alpha=0$. The compressed-exponential tail of $\varrho(w)$ derived above (exponential for $\alpha=0$) ensures that the integral $\int |x|^\beta P_{\mathrm{st}}(x)\,dx$ converges for any $\beta>0$ and any $\alpha\geq0$, so that the cost rate is finite at every $r>0$. The effect of sluggishness appears in the weak-resetting limit $r\to0$. Equation~\eqref{eq:avgcost_pow} shows that the cost rate then vanishes for $\beta<\alpha+2$, tends to a constant for $\beta=\alpha+2$, and diverges for $\beta>\alpha+2$. For normal diffusion the cost rate diverges as $r\to0$ for $\beta>2$~\citep{Sunil_2023}, and sluggish dynamics shifts the critical exponent from $2$ to $\alpha+2$, see Sec.~\ref{sec:poisson_zero_r_limit}.

\begin{figure}[h]
    \centering
    \includegraphics[width=0.75\textwidth, trim=0cm 0cm 0cm 0cm, clip]{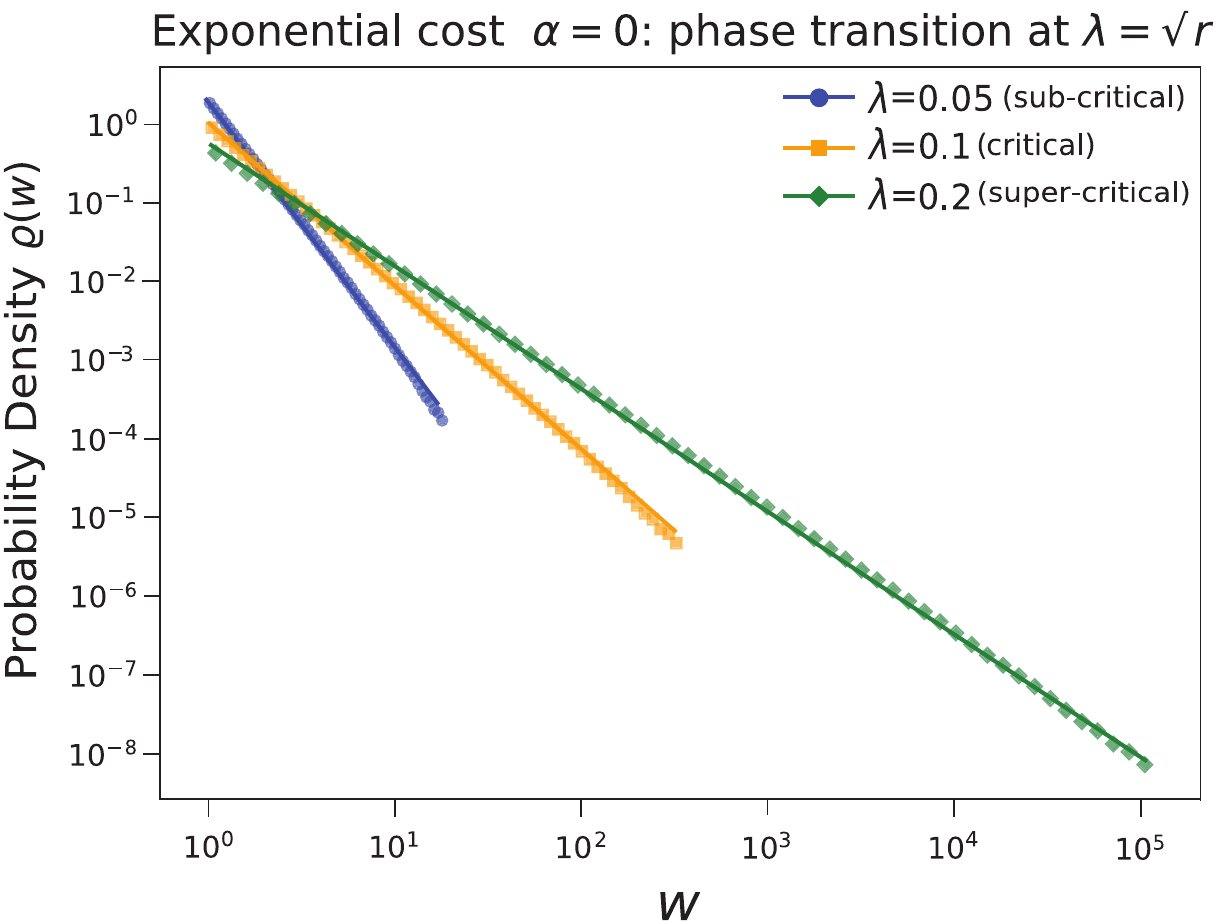}
    \caption{Probability density $\varrho(w)$ of the cost per reset for exponential cost $c(|x|)=e^{\lambda|x|}$ at $\alpha=0$ under Poissonian resetting, for resetting rate $r=0.01$ and $\lambda \in \{0.05,\, 0.1,\, 0.2\}$ spanning the sub-critical, critical, and super-critical regimes about $\lambda=\sqrt{r}$ (see legend). Symbols are obtained from $N = 10^6$ independent simulations, each evolved until its first reset, with the cost evaluated at the position just before the reset. Lines are given by the exact closed form $\varrho(w)=(\sqrt{r}/\lambda)\,w^{-1-\sqrt{r}/\lambda}$.}
\label{fig:9}
\end{figure}

\paragraph{Exponential cost.}
Setting $c(|x|) = e^{\lambda|x|}$ with $\lambda > 0$ models situations where long-distance returns are heavily penalised~\citep{Sunil_2023}. The tail of the integrand then behaves as
\begin{equation}
    e^{\lambda|x|} P_{\mathrm{st}}(x)
        \sim \exp\!\left(\lambda|x|
             - \frac{2\sqrt{r}}{\alpha+2}|x|^{(\alpha+2)/2}\right),
    \label{eq:exp_competition}
\end{equation}
and convergence is determined by the competition between the two exponentials in~\eqref{eq:exp_competition}, consistent with the power-law tail of $\varrho(w)$ found above for $\alpha=0$. For $\alpha > 0$, the decay exponent $|x|^{(\alpha+2)/2}$ grows faster than linearly, so the integral converges for all $\lambda > 0$ and the cost rate is finite at any $r>0$. The exponent in~\eqref{eq:exp_competition} is maximal at $x_c=(\lambda/\sqrt{r})^{2/\alpha}$, where it reaches $\alpha\lambda x_c/(\alpha+2)$. For $\alpha = 0$ the two terms compete linearly and the well-known phase transition at $\lambda = \sqrt{r}$ is recovered~\citep{Sunil_2023}, as shown in Fig.~\ref{fig:9}, where $\varrho(w)\sim w^{-1-\sqrt{r}/\lambda}$ has a finite mean only for $\lambda<\sqrt{r}$, and its power-law tail flattens as $\lambda$ increases towards $\sqrt{r}$. The mean cost rate of the exponential cost is non-monotonic in $r$. It grows without bound as $r\to\lambda^2$ from above (at $\alpha=0$) or as $r\to0$ (at $\alpha>0$), and grows linearly with $r$ at large $r$, since the walker is then reset before it moves away from the origin and $c\to1$. It therefore has an interior minimum. For $\alpha=0$, Eq.~\eqref{eq:avgcost} gives $\langle\dot{\mathcal{C}}\rangle_{\rm st}=r^{3/2}/(\sqrt{r}-\lambda)$ for $r>\lambda^2$, minimised at $r^*=9\lambda^2/4$ with minimum value $27\lambda^2/4$. For $\alpha>0$ the minimum is found numerically and moves to smaller $r$ as $\alpha$ increases. For $\lambda=0.05$ it lies at $r^*\simeq5.6\times10^{-3}$, $7.6\times10^{-4}$ and $9\times10^{-6}$ for $\alpha=0$, $0.5$ and $1.5$, with minimal cost rates of about $1.7\times10^{-2}$, $3.9\times10^{-3}$ and $1.3\times10^{-4}$, respectively. In contrast, the linear and power-law costs give pure power laws in $r$, Eqs.~\eqref{eq:avgcost_lin} and~\eqref{eq:avgcost_pow}, which are monotonic and have no optimal rate.

More general costs of the form $c(|x|)=e^{\lambda_0|x|^{\gamma}}$ interpolate between these cases. The convergence of the integral is again controlled by the competition between $\lambda_0|x|^{\gamma}$ and the decay exponent $\frac{2\sqrt{r}}{\alpha+2}|x|^{(\alpha+2)/2}$ of Eq.~\eqref{eq:asympt}. For $\gamma<(\alpha+2)/2$ the decay always wins and the cost rate is finite for any $\lambda_0$, whereas for $\gamma>(\alpha+2)/2$ the cost wins and the cost rate diverges for any $\lambda_0>0$. At the marginal value $\gamma=(\alpha+2)/2$ the exponentials cancel when $\lambda_0=2\sqrt{r}/(\alpha+2)$, leaving an integrand that does not decay, since it behaves as $|x|^{3\alpha/4}$, and finiteness therefore requires the strict inequality $\lambda_0<2\sqrt{r}/(\alpha+2)$. The exponential cost corresponds to $\gamma=1$, which is marginal only at $\alpha=0$. These results are summarised in Table~\ref{tab:phases}.

\begin{table}
\renewcommand{\arraystretch}{1.6}
\caption{Phase structure of the mean cost rate $\langle\dot{\mathcal{C}}\rangle_{\mathrm{st}}$ for different cost functions under Poissonian resetting. Finiteness refers to a fixed $r>0$ unless the limit $r\to0$ is stated. The sluggishness exponent $\alpha$ qualitatively changes the convergence criteria compared to normal diffusion ($\alpha = 0$).}
\centering
\begin{tabular}{| l | c | c |}
\hline
\textbf{Cost function} & \textbf{Condition for finite
$\langle\dot{\mathcal{C}}\rangle_{\mathrm{st}}$} & \textbf{Role of $\alpha$} \\
\hline
$|x|^\beta$, $\beta > 0$ &
\makecell{finite for all $r>0$ and $\alpha\geq0$, \quad~\eqref{eq:avgcost_pow}\\[4pt]
as $r\to0$, $\langle\dot{\mathcal{C}}\rangle_{\mathrm{st}}\sim r^{1-\beta/(\alpha+2)}$\\[4pt]
vanishes for $\beta<\alpha+2$,\\[4pt]
diverges for $\beta>\alpha+2$} &
\makecell{shifts the critical\\ exponent from $\beta=2$\\ to $\beta=\alpha+2$} \\
\hline
\multirow{2}{*}{$e^{\lambda|x|}$}
  & $\alpha = 0$: $\lambda < \sqrt{r}$, \qquad~\eqref{eq:exp_competition} & $\alpha$ determines existence \\
  & $\alpha > 0$: finite for all $\lambda$, \qquad~\eqref{eq:exp_competition} &  \\
\hline
\multirow{3}{*}{$e^{\lambda_0 |x|^\gamma}$}
  & $\gamma < \frac{\alpha+2}{2}$: finite, \qquad~\eqref{eq:asympt} &  \\
  & $\gamma > \frac{\alpha+2}{2}$: divergent & $\alpha$ sets critical exponent \\
  & $\gamma = \frac{\alpha+2}{2}$: finite iff $\lambda_0 < \frac{2\sqrt{r}}{\alpha+2}$ &  \\
\hline
\end{tabular}
\label{tab:phases}
\end{table}

\subsubsection{Scaling in the zero-resetting-rate limit} \label{sec:poisson_zero_r_limit}
The limit $r \to 0$ is physically instructive as it reveals how the cost accumulates when resets are rare but potentially very costly~\citep{Sunil_2023}. The exact result~\eqref{eq:avgcost_pow} admits a simple physical interpretation. As $r \to 0$, the typical distance a particle diffuses between successive resets is set by the length scale obtained by inverting the compressed-exponential decay in~\eqref{eq:asympt}, giving $\ell \sim r^{-1/(\alpha+2)}$. For a power-law cost $c(|x|) \sim |x|^\beta$, the average cost incurred in a single reset therefore scales as $\langle c \rangle \sim \ell^\beta \sim r^{-\beta/(\alpha+2)}$. Since the mean number of resets up to time $t$ is $\langle N{(t)} \rangle \sim rt$, the total expected cost behaves as
\begin{equation}
    \langle \mathcal{C}(t) \rangle
        \sim \langle N(t) \rangle\,\langle c \rangle
        \sim r^{1 - \beta/(\alpha+2)}\,t,
    \label{eq:scaling}
\end{equation}
which reproduces the $r$-dependence of the exact result~\eqref{eq:avgcost_pow}. The mean cost rate $\langle\dot{\mathcal{C}}\rangle = \lim_{t\to\infty} \langle \mathcal{C}(t)\rangle/t$ therefore exhibits three distinct regimes as $r \to 0$.
\begin{enumerate}
    \item if $\beta < \alpha+2$, the exponent is positive and the rate vanishes,
    \item if $\beta = \alpha+2$, the rate approaches a finite constant,
    \item if $\beta > \alpha+2$, the rate diverges.
\end{enumerate}
For normal diffusion ($\alpha = 0$), the critical exponent is $\beta = 2$~\citep{Sunil_2023}. For $\alpha > 0$, the threshold shifts to $\beta = \alpha + 2 > 2$, so a wider class of super-linear costs keeps a bounded cost rate in the weak-resetting limit. This reflects the physical fact that sluggish dynamics suppresses large excursions, making the rare but costly resets less probable than in the normal-diffusion case.

\subsection{Threshold Resetting case}\label{sec:threshold_cost}
\smallskip
Threshold resetting is purely position-driven with a reset occurring whenever the walker reaches $|x| = x_{\mathrm{th}}$, making each reset cycle a first-passage event~\citep{Redner_2001, Bhat_2016}. As a consequence, there is no resetting rate $r$ to optimise and the natural control parameter is the threshold distance $x_{\mathrm{th}}$. Every reset also occurs from the same distance, so that the cost per reset is deterministic and equal to $c(x_{\mathrm{th}})$. Since each cycle starts from the origin and has the mean duration $\langle\tau\rangle=x_{\rm th}^{\alpha+2}/[(\alpha+1)(\alpha+2)]$, see Eq.~\eqref{eq:cycle_time}, the cost rate is
\begin{equation}
    \langle\dot{\mathcal{C}}\rangle_{\rm th}
        = \frac{c(x_{\rm th})}{\langle\tau\rangle}
        = \frac{(\alpha+1)(\alpha+2)\,c(x_{\rm th})}{x_{\rm th}^{\alpha+2}},
    \label{eq:cost_rate_thresh}
\end{equation}
which is finite for any finite $x_{\mathrm{th}}$, in contrast to the Poissonian case where divergences can arise at finite $r$~\citep{Sunil_2023}. The interesting question is instead how the cost rate scales as $x_{\mathrm{th}} \to \infty$, when the confinement weakens, and the walker can wander arbitrarily far before being reset. This is addressed at the end of this subsection.

Under Poissonian resetting, the position from which a reset occurs is distributed according to the stationary distribution. This is not the case for threshold resetting. The stationary distribution~\eqref{eqn:posfeed} has compact support in $(-x_{\mathrm{th}}, x_{\mathrm{th}})$ and describes the position of the walker at a random time rather than at reset events. Weighted by a cost function, it gives the distribution of the instantaneous cost $w=c(|x|)$ at a random time, such as a running cost paid per unit time while the walker is at $x$, which is bounded above by $c(x_{\mathrm{th}})$. As in Sec.~\ref{sec:poisson_cost}, we first derive the distribution of this stationary cost, obtain its mean, and then examine the large-threshold scaling of the cost rate, for linear, power-law, and exponential cost functions.

\subsubsection{Distribution of the stationary cost}\label{sec:threshold_cost_dist}
The full distribution of the stationary cost $w=c(|x|)$ is obtained by the same change-of-variables argument used in Sec.~\ref{sec:poisson_cost_dist}. In the steady state, the position of the walker at a random time is distributed according to $P_{\mathrm{st}}(x)$ given in~\eqref{eqn:posfeed}, so that
\begin{equation}
    \varrho(w) = 2\, P_{\mathrm{st}}\!\bigl(x(w)\bigr)
           \left|\frac{dx}{dw}\right|,
    \qquad w_{\min} \leq w \leq w_{\max},
    \label{eq:costdist_thresh}
\end{equation}
where the upper limit $w_{\max} = c(x_{\mathrm{th}})$ is finite because the support of $P_{\mathrm{st}}$ is compact. The stationary cost distribution has a hard upper cutoff, so no heavy tails or divergent moments can arise for any finite $x_{\mathrm{th}}$, as we now show for the same three cost functions considered above.

\begin{itemize}
        \item {\bf Linear cost.}
For $w = |x|$, the distribution is proportional to $P_{\mathrm{st}}$ yielding,
\begin{equation}
    \varrho(w) = \frac{(\alpha+1)(\alpha+2)}{x_{\mathrm{th}}^{\alpha+2}}\,
           w^\alpha (x_{\mathrm{th}} - w),
    \qquad 0 \leq w \leq x_{\mathrm{th}}.
    \label{eq:fw_lin_thresh}
\end{equation}
Equation~\eqref{eq:fw_lin_thresh} is a Beta distribution on $[0, x_{\mathrm{th}}]$ with shape parameters $\alpha+1$ and $2$. For $\alpha = 0$ the distribution is linear in $w$ with a negative slope, while for $\alpha > 0$ it develops a peak at $w^* = \alpha x_{\mathrm{th}} / (\alpha+1)$ that moves towards $x_{\mathrm{th}}$ as $\alpha$ increases, reflecting the tendency of the sluggish walker to spend more time far from the origin before being reset. Near the upper boundary, $\varrho(w) \sim (x_{\mathrm{th}} - w)$, so the distribution vanishes linearly rather than exhibiting fat tails.

\item {\bf Power-law cost.}
For $w = |x|^\beta$, one has $x = w^{1/\beta}$ and $dx/dw = \beta^{-1} w^{1/\beta - 1}$, giving
\begin{equation}
    \varrho(w) = \frac{(\alpha+1)(\alpha+2)}{\beta\, x_{\mathrm{th}}^{\alpha+2}}\,
           w^{\frac{\alpha+1}{\beta} - 1}
           \left(x_{\mathrm{th}} - w^{1/\beta}\right),
    \qquad 0 \leq w \leq x_{\mathrm{th}}^\beta.
    \label{eq:fw_power_thresh}
\end{equation}
The factor $(x_{\mathrm{th}} - w^{1/\beta})$ provides a soft linear cutoff near the maximum cost $w_{\max} = x_{\mathrm{th}}^\beta$, and all moments of $\varrho(w)$ are finite for any $\alpha \geq 0$.

\item {\bf Exponential cost.}
For $w = e^{\lambda|x|}$, one has $|x| = \lambda^{-1} \ln w$ and $dx/dw =
1/(\lambda w)$, leading to
\begin{equation}
    \varrho(w) = \frac{(\alpha+1)(\alpha+2)}{\lambda^{\alpha+1}\, x_{\mathrm{th}}^{\alpha+2}}\,
           \frac{(\ln w)^\alpha}{w}
           \left(x_{\mathrm{th}} - \frac{\ln w}{\lambda}\right),
    \qquad 1 \leq w \leq e^{\lambda x_{\mathrm{th}}}.
    \label{eq:fw_exp_thresh}
\end{equation}
The distribution is supported on the finite interval $[1,\, e^{\lambda x_{\mathrm{th}}}]$ and vanishes at the upper boundary, so all moments are finite.
\end{itemize}

\begin{figure}
    \centering
    \includegraphics[width=1\textwidth, trim=0cm 0cm 0cm 0cm, clip]{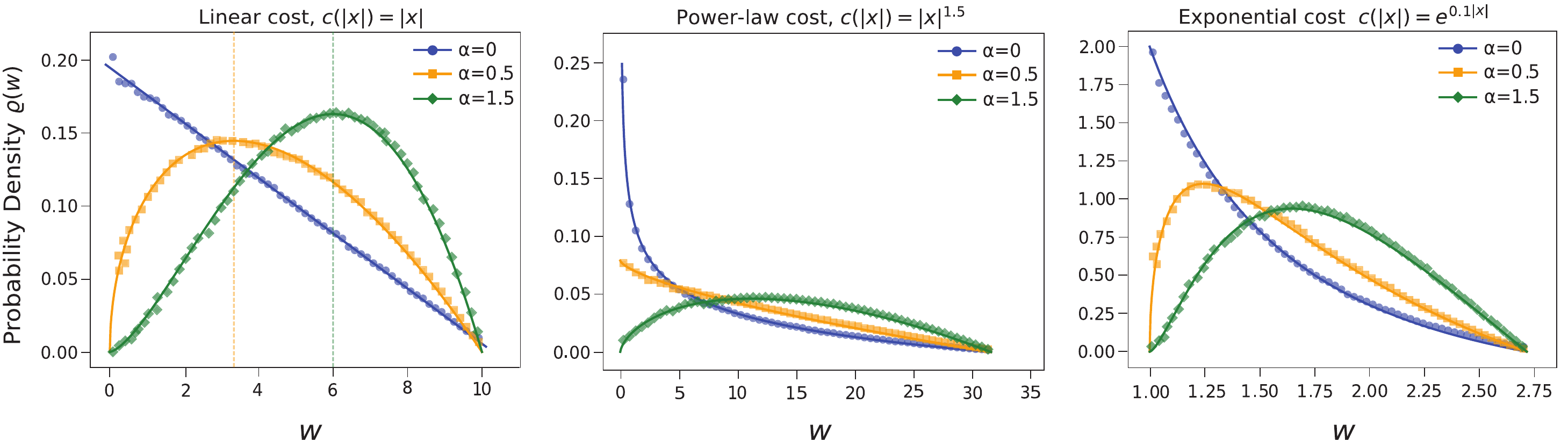}
    \caption{Stationary probability density $\varrho(w)$ of the instantaneous cost $w=c(|x|)$ under threshold resetting, for linear (left panel), power-law with $\beta=1.5$ (middle panel), and exponential with $\lambda=0.1$ (right panel) cost functions, and sluggishness parameters $\alpha \in \{0,\, 0.5,\, 1.5\}$ (see legend), at fixed $x_{\rm th}=10$. Symbols are obtained from $N = 10^6$ independent simulations evolved to $T=10\langle\tau\rangle$~\eqref{eq:cycle_time}, which ensures stationarity for each $\alpha$, with the cost evaluated at the final position. Lines are given by the exact analytical results~\eqref{eq:fw_lin_thresh}--\eqref{eq:fw_exp_thresh}. Dashed vertical lines in the left panel mark the peak location $w^*=\alpha x_{\rm th}/(\alpha+1)$ for $\alpha>0.$}
\label{fig:10}
\end{figure}

The compact support imposed by threshold resetting suppresses fat tails and diverging mean costs, regardless of the value of $\alpha$, as shown in Fig.~\ref{fig:10} where all three cost distributions terminate sharply at a finite $w_{\max}$.

\subsubsection{Average stationary cost}\label{sec:threshold_avg_cost}
The average stationary cost follows as the first moment of the distribution derived above,
\begin{equation}
    \langle c \rangle_{\mathrm{st}}
        = \int_{-x_{\mathrm{th}}}^{x_{\mathrm{th}}}
          c(|x|)\, P_{\mathrm{st}}(x)\, dx
        = \int_{w_{\min}}^{w_{\max}} w\, \varrho(w)\, dw.
    \label{eq:avgcost_thresh}
\end{equation}
Because $P_{\mathrm{st}}$ has compact support, this integral reduces to a definite integral over a finite domain and can be evaluated in closed form for several cost functions of interest. This average of $c(|x|)$ over the stationary distribution equals the mean cost rate of a running cost paid per unit time at position $x$. It should not be confused with the cost per reset $c(x_{\rm th})$ and the cost rate~\eqref{eq:cost_rate_thresh} discussed above.

\begin{itemize}
    \item {\bf Linear cost.}
Setting $c(|x|) = |x|$ and substituting~\eqref{eqn:posfeed} into~\eqref{eq:avgcost_thresh} gives
\begin{equation}
    \langle c \rangle_{\mathrm{lin}}
        = \frac{(\alpha+1)(\alpha+2)}{x_{\mathrm{th}}^{\alpha+2}}
          \int_0^{x_{\mathrm{th}}} x^{\alpha+1}(x_{\mathrm{th}} - x)\, dx
        = \frac{\alpha+1}{\alpha+3}\, x_{\mathrm{th}}.
    \label{eq:avg_lin_thresh}
\end{equation}
For $\alpha = 0$, this gives $\langle c \rangle_{\mathrm{lin}} =x_{\mathrm{th}}/3$, while for $\alpha > 0$ the prefactor $(\alpha+1)/(\alpha+3) < 1$ increases with increasing sluggishness, from $1/3$ at $\alpha=0$ towards $1$ for large $\alpha$, reflecting the fact that the stationary distribution~\eqref{eqn:posfeed} shifts probability towards the boundary at $|x| = x_{\mathrm{th}}$, with its peak at $\alpha x_{\mathrm{th}}/(\alpha+1)$. In both cases, the mean cost grows linearly with the threshold value.

\item {\bf Power-law cost.}
For $c(|x|) = |x|^\beta$ with $\beta > 0$, the integral is again elementary and yields the exact result
\begin{equation}
    \langle c \rangle_{\beta}
        = \frac{(\alpha+1)(\alpha+2)}{(\alpha+\beta+1)(\alpha+\beta+2)}\,
          x_{\mathrm{th}}^\beta.
    \label{eq:avg_power_thresh}
\end{equation}
The mean cost scales as $x_{\mathrm{th}}^\beta$ for all $\beta > 0$ and all $\alpha \geq 0$, with a prefactor that increases monotonically with $\alpha$ for fixed $\beta$ (for $\beta=1.5$ it equals $0.23$, $0.31$ and $0.44$ for $\alpha=0$, $0.5$ and $1.5$). Sluggish dynamics therefore increases the mean stationary cost at any given threshold, since the stationary distribution shifts weight towards larger displacements, where the walker lingers. 

\item {\bf Exponential cost.}
For $c(|x|) = e^{\lambda|x|}$ with $\lambda > 0$, direct integration for $\alpha = 0$ gives the exact result
\begin{equation}
    \langle c \rangle_{\mathrm{exp}}\big|_{\alpha=0}
        = \frac{2}{(\lambda x_{\mathrm{th}})^2}
          \left(e^{\lambda x_{\mathrm{th}}} - \lambda x_{\mathrm{th}} - 1\right),
    \label{eq:avg_exp_thresh_alpha0}
\end{equation}
which is manifestly positive for all $\lambda x_{\mathrm{th}} > 0$, since $e^z > 1+z$. For general $\alpha$, the integral $\int_0^{x_{\mathrm{th}}} x^\alpha e^{\lambda x}(x_{\mathrm{th}}-x)\,dx$ is not needed in closed form here, since the large-$x_{\mathrm{th}}$ behaviour follows from a Laplace-type argument: the integral is dominated by the boundary layer $x = x_{\mathrm{th}} - s$ with $s = O(1/\lambda) \ll x_{\mathrm{th}}$, where $x^\alpha \approx x_{\mathrm{th}}^\alpha$ and $e^{\lambda x} \approx e^{\lambda x_{\mathrm{th}}}e^{-\lambda s}$, giving
\begin{equation}
    \langle c \rangle_{\mathrm{exp}}
        \sim \frac{(\alpha+1)(\alpha+2)}{\lambda^2 x_{\mathrm{th}}^2}\,
             e^{\lambda x_{\mathrm{th}}},
    \qquad x_{\mathrm{th}} \to \infty,
    \label{eq:avg_exp_thresh}
\end{equation}
which recovers the leading large-$x_{\mathrm{th}}$ behaviour of the exact result~\eqref{eq:avg_exp_thresh_alpha0} at $\alpha=0$. This asymptotic form is controlled by the region near $x = x_{\mathrm{th}}$, where the exponentially growing cost $c(x) = e^{\lambda|x|}$ outweighs the linear vanishing of the stationary density~\eqref{eqn:posfeed}.
\end{itemize}

\subsubsection{Large-threshold scaling and effective phase transition}
Although the cost per reset is always finite for finite $x_{\mathrm{th}}$, the behaviour of the cost rate~\eqref{eq:cost_rate_thresh} as $x_{\mathrm{th}} \to \infty$ reveals a rich phase structure analogous to the zero-resetting-rate limit of Poissonian resetting (Sec.~\ref{sec:poisson_zero_r_limit}). For a power-law cost $c(|x|) = |x|^\beta$ one has $\langle\dot{\mathcal{C}}\rangle_{\rm th}=(\alpha+1)(\alpha+2)\,x_{\rm th}^{\beta-(\alpha+2)}$, where the factor $x_{\rm th}^{\alpha+2}$ comes from the mean cycle duration. One thus identifies three distinct regimes for the cost rate.
\begin{enumerate}
    \item if $\beta < \alpha+2$, the cost rate $\langle\dot{\mathcal{C}}\rangle_{\rm th}\propto x_{\mathrm{th}}^{\beta-(\alpha+2)} \to 0$ as
    $x_{\mathrm{th}} \to \infty$,
    \item if $\beta = \alpha+2$, it approaches the constant $(\alpha+1)(\alpha+2)$,
    \item if $\beta > \alpha+2$, it diverges as $x_{\mathrm{th}}^{\beta-(\alpha+2)}$.
\end{enumerate}
The sluggishness exponent $\alpha$ therefore sets the critical value $\beta_c=\alpha+2$, as in the Poissonian case (Sec.~\ref{sec:poisson_zero_r_limit} and summarised in Table~\ref{tab:phases}), with $x_{\mathrm{th}}$ playing the role of the characteristic diffusion length $\ell \sim r^{-1/(\alpha+2)}$ and the mean cycle duration $\langle\tau\rangle\propto x_{\mathrm{th}}^{\alpha+2}$ playing the role of $1/r$~\citep{Sunil_2023}. These scaling regimes are summarised in Table~\ref{tab:scaling_thresh}.

\begin{table}
\caption{Large-threshold scaling regimes of the cost rate $\langle\dot{\mathcal{C}}\rangle_{\rm th}=c(x_{\rm th})/\langle\tau\rangle$ for a power-law cost $c(|x|) = |x|^\beta$ under threshold resetting. The sluggishness exponent $\alpha$ sets the critical value $\beta_c = \alpha+2$, in direct analogy with the $r \to 0$ scaling of Poissonian resetting.}
\centering
\renewcommand{\arraystretch}{1.6}
\begin{tabular}{|l | c|}
\hline
\textbf{Condition} &
\textbf{Scaling of $\langle\dot{\mathcal{C}}\rangle_{\rm th}$ as $x_{\mathrm{th}} \to \infty$} \\
\hline
$\beta < \alpha+2$ & $\sim x_{\mathrm{th}}^{\beta-(\alpha+2)} \to 0$ \\
$\beta = \alpha+2$ & constant, $(\alpha+1)(\alpha+2)$ \\
$\beta > \alpha+2$ & $\sim x_{\mathrm{th}}^{\beta-(\alpha+2)} \to \infty$ \\
\hline
\end{tabular}
\label{tab:scaling_thresh}
\end{table}

\section{Conclusion}
\label{sec:8}
In this paper, we have introduced and solved exactly two models coupling sluggish diffusion to two paradigmatic resetting protocols: Poissonian resetting at a constant rate $r$, and threshold resetting triggered upon reaching a fixed distance $x_{\rm th}$. Our results demonstrate that the sluggishness parameter $\alpha$ does not merely rescale the known phenomenology of normal diffusion with resetting; it qualitatively reshapes the nonequilibrium stationary state, the first-passage and occupation statistics, and the cost of sustaining the process out of equilibrium.

Under Poissonian resetting, we derived the exact stationary distribution~(\ref{eqn:bessel}) in terms of a modified Bessel function, which collapses onto an $r$-independent curve in the variable $x\,r^{1/(\alpha+2)}$. It reduces to the bilateral exponential of normal diffusion at $\alpha=0$ and develops an increasingly  sharp bimodal, cusped structure as $\alpha$ increases, reflecting the competition between the sluggish dynamics and the resetting current. Under threshold resetting, the stationary distribution is an exact Beta-type law (\ref{eqn:posfeedscale}) whose mode migrates continuously from the origin towards the confining boundary as $\alpha$ increases. In both cases, resetting converts an otherwise divergent mean first-passage time, a consequence of the heavy power-law tail $F_0 \sim t^{-1-\mu}$ with $\mu = 1/(\alpha+2)$, into a finite one. Under Poissonian resetting the mean first-passage time has an optimal resetting rate, $r^*$, with $r^*x_0^{\alpha+2}$ a function of $\alpha$ only, whereas under threshold resetting the mean first-passage time to the origin increases monotonically with the threshold. The occupation fraction follows the Lamperti law~(\ref{eq:f_plus}) for the free walk and for $rT\ll1$ under Poissonian resetting, becomes a Gaussian centred at $1/2$ with variance $(1-\mu)/(2rT)$ for $rT\gg1$, and converges to the deterministic value $\xi^{*} \in (1/2, 1-\mu/2)$~(\ref{eq:xi_star_closed}) under threshold resetting.

For the cost of resetting, the analysis separates two notions of cost. The resetting entropy production rate $\dot S_{\rm reset}$ equals $-r$ at $\alpha=0$ but diverges in the continuum for every $\alpha>0$, because the stationary density vanishes at the reset point and resetting into it carries a diverging informational cost; on the lattice it is finite and grows only logarithmically with the ratio $\ell/b$ between the length scale of the stationary state and the lattice spacing. The phenomenological cost per reset $c(|x|)$, a resource penalty such as the time or energy needed to return the walker, remains well defined for $\alpha>0$. Under Poissonian resetting, the compressed-exponential tail of the stationary distribution keeps the mean cost per reset finite at any $r>0$ for power-law costs, and the mean cost rate scales as $r^{1-\beta/(\alpha+2)}$ as $r\to0$, so that the critical exponent moves from $\beta=2$ at $\alpha=0$ to $\beta_c=\alpha+2$, and a wider class of super-linear costs keeps a bounded cost rate. For the exponential cost, the mean cost rate diverges for $\lambda\geq\sqrt{r}$ at $\alpha=0$, is finite for every $\lambda>0$ at $\alpha>0$, and has an interior minimum as a function of $r$. Under threshold resetting the cost per reset is fixed to $c(x_{\rm th})$ and is finite for any $x_{\rm th}$, and the same $\beta_c=\alpha+2$ governs the large-threshold scaling of the cost rate. Sluggishness thus acts in opposite directions on the two notions of cost: it makes the informational cost of resetting diverge in the continuum, while it raises the critical exponent of the resource cost.

These results open several directions for future work. Extending this framework to higher dimensions would clarify how the sluggishness exponent interacts with resetting geometry beyond the one-dimensional bimodal and boundary-peaked profiles obtained here. Resetting to a non-zero or randomly drawn point in the stationary state would break the symmetry exploited for the stationary distributions in this work and is expected to couple $\alpha$ and the resetting protocol in new ways. A further open question is the resetting entropy production for sluggish dynamics in the continuum, where the singular effective potential at the origin makes the stationary density vanish at the reset point and calls for a regularised or coarse-grained treatment of the resetting-point entropy. Finally, the cost results motivate a full optimal-control treatment that jointly optimises the resetting rate or threshold against both search-time and energetic objectives, extending the mean-completion-time framework of \cite{Sunil_2023, Pal_2020} to the sluggish setting studied here.

%





\bibliography{references}

\end{document}